\documentclass[]{article}
\usepackage{amsmath,amssymb,amsthm}
\title{Lee model and its  Resolvent Analysis}
\usepackage{authblk}
\usepackage{titlesec}
\usepackage[toc,page]{appendix}
\author[1]{Yesukhei Jagvaral}
\author[1,2]{O. Teoman Turgut \footnote{corresponding author: turgutte@boun.edu.tr}}
\author[3]{ Meltem \"Unel}

\affil[1]{\footnotesize \textit{Department of Physics, Carnegie Mellon University, Pittsburgh, PA 15213, USA }}
\affil[2]{\footnotesize \textit{Department of Physics, Bo˘gazi¸ci University, Bebek, 34342, ˙Istanbul, Turkey }}
\affil[3]{\footnotesize \textit{Department of Mathematical Sciences, University of Copenhagen, DK-2100, Copenhagen, Denmark}}

\date{\vspace{-5ex}}

\usepackage{color}
\usepackage{indentfirst}
\usepackage{multirow}
\usepackage{subfigure}
\usepackage{algorithm}
\usepackage{algorithmic}
\usepackage[bottom]{footmisc}
\usepackage{cite}
\usepackage{multicol}
\usepackage{graphicx}
\usepackage{longtable}
\usepackage{mathtools}
\usepackage{mathrsfs}
\usepackage{epigraph}
\DeclarePairedDelimiter\bra{\langle}{\rvert}
\DeclarePairedDelimiter\ket{\lvert}{\rangle}
\DeclarePairedDelimiterX\braket[2]{\langle}{\rangle}{#1 \delimsize\vert #2}

\newcommand{\xt}{{x}^\bot}

\newcommand{\pt}{{p}_\bot}
\newcommand{\qt}{{q}_\bot}
\newcommand{\lip}{\int_0^\infty \frac{d p}{2 \pi} \,}
\newcommand{\liq}{\int_0^\infty \frac{d q}{2 \pi} \,}
\newcommand{\ipt}{\int \frac{d p_\bot}{2 \pi} \,}
\newcommand{\iqt}{\int \frac{d q_\bot}{2 \pi} \,}
\newcommand{\lop}{\omega(p,\pt)}
\newcommand{\loq}{\omega(q,\qt)}

\newcommand{\lap}{a(p,\pt)}

\newcommand{\ladp}{a^\dag(p,\pt)}
\newcommand{\ladq}{a^\dag(q,\qt)}

\newtheorem{thm}{Theorem}
\newtheorem{definition}{Definition}
\newtheorem*{thm*}{Theorem}
\newtheorem*{cor*}{Theorem}
\newtheorem*{def*}{Definition}

\newcommand{\bb}{\begin{equation}}
\newcommand{\ee}{\end{equation}}
\newcommand{\p}{\partial}

\newcommand*{\defeq}{\mathrel{\vcenter{\baselineskip0.5ex \lineskiplimit0pt
			\hbox{\scriptsize.}\hbox{\scriptsize.}}}%
	=}

\begin{document}
\maketitle

\begin{abstract}
We revisit the relativistic 2+1 dimensional Lee model on flat space in light-front coordinates  and  on a  space-time with a spatial section given by a  compact manifold, in  the usual canonical  formalism. The simpler 2+1 dimension is chosen because  renormalization is needed only for the mass difference but not required for  the coupling constant and the wavefunction. The model is constructed non-perturbatively based on the  resolvent formulation \cite{kaynak2009relativistic}. The bound state spectrum is studied through its  ``principal operator" and bounds for the ground state energy are obtained. We show that the formal expression found indeed defines the resolvent of a self-adjoint operator--the Hamiltonian of the interacting system.  Moreover, we prove an essential result that the principal operator corresponds to a self-adjoint holomorphic family of type-A, in the sense of Kato. 
\end{abstract}

 \section*{Introduction}
 
 Lee model is a nontrivial toy model originally proposed for the purpose of understanding renormalization in a non-perturbative way. The model contains two fermion species, $V$ and $N$ particles. Although they are allowed to carry nontrivial momenta, their energies are assumed to be given by their masses as they are considered to be heavy particles \cite{lee1954some, pauli-kallen}. They interact with a relativistic real scalar field, the $\theta$ particle,  where the only interaction allowed is:
 \begin{equation}
 N+\theta\leftrightarrow V    
 \end{equation}
 i.e. no crossing symmetry is present. The elimination of the crossing symmetry is possible by truncating the field operators of $\theta$ to positive and negative frequencies respectively. For a detailed discussion of the original model, we refer to the existing literature  \cite{schweber2011introduction,decelles1959,amado, sommerfield1965,bolsterli1968,pagnamenta1965, maxon1965, bolsterli1983, Fuda1982}. It must be noted that some of these works focus on a simplified version where it is assumed that there is no momentum transfer between the fermions and the bosons. 
Since the interaction term leads to the conservation of total fermion number, one can restrict the Hamiltonian to the subspace corresponding to a single $V$ or $N$ particle together with an arbitrary number of bosons. Hence, in the aforementioned subspace, ignoring the recoil, we can simplify the model as a {\it fixed} two level system interacting with an arbitrary number of bosons. This is the setup that we focus on in this work and from this point on, it will be called \textit{Restricted Lee Model} (RLM) (a version of which is also known as  the Wigner-Weisskopf model). 
This restriction was also employed by Wilson \cite{wilson1970model} in his analysis of Lee model with crossing symmetry where he studied coupling constant renormalization in a non-perturbative manner. 
A Poincare invariant version of the Lee model is constructed in light-front coordinates in \cite{Fuda1990} and many subtleties about the relativistic invariance are discussed therein. 

Thirring and Henley studies RLM with a further simplification where the bosons are treated by a non-relativistic dispersion relation \cite{thirring-book}. In an unpublished inspiring work Rajeev applies an algebraic approach to obtain the resolvent of this very version \cite{rajeev1999bound}. His construction naturally leads to an operator, \textit{the Principal Operator}, which contains all the information about the bound states. Rajeev uses some estimates to guarantee invertibility of  the principal operator to find a lower bound for the ground state, for an arbitrary number of bosons.
In this work, we study relativistic RLM on two different backgrounds: on flat space in light-front coordinates  and  on a  space-time with the spatial sections given by a compact manifold. Following Rajeev's approach we aim to justify the formal manipulations that are used to obtain most of the aforementioned results and indeed show that there is a well-defined Hamiltonian underlying the obtained resolvent. 
To establish this, we use some ideas about pseudo-resolvents from semi-group theory, instead of the more widely used resolvent convergence techniques. We define the theory via the resolvent and in this case, it does not matter what limit process is used in its construction. Having found the resolvent, one needs to check the self-adjointness and as a result, compute the ground state wave function via a spectral projection. The validity of this computation requires the Principal Operator, as a function of the complex parameter $E$, to be a self-adjoint holomorphic family of type-A in the sense of Kato. It is one of the main results of this work to verify this claim. This  approach was  applied  to the non-relativistic Lee model in a previous work \cite{teo-erman-dogan}. This approach has the following advantage, in some cases a cut-off regularization is not the most convenient one and  the regularized expressions may not correspond to the resolvent of an operator at this intermediate stage, although the limit may be a sensible resolvent. Having completed this work we became aware of the articles by Arai and Hirokawa in which    Wigner-Weisskopf model with a more general dispersion relation for bosons and assuming smeared out couplings to the two level system  are studied by operator theory methods in the lowest particle sectors  \cite{hirokawa1, hirokawa2}. In a recent work by Facchi et al \cite{facchi}, the spectral properties of the  RLM is revisited rigorously by employing slightly different regularization methods.

 Due to the effective nature of Quantum Field Theory, one expects divergences that must be removed by redefinition of some finite number of parameters. It is essential to verify that the resulting theory, after the removal of all divergences, is mathematically meaningful. Up to the present time, this has not been possible in realistic theories, despite the efforts of many people - one possible starting point could be the work of Glimm and Jaffe \cite{glimm-jaffe}.
 Our work is a modest attempt in this direction for a very simple model. Although the discussion is highly technical from a theoretical physics perspective, not  complete mathematical rigor is aimed since the purpose is rather to establish the essential ideas.

 \subsection*{Outline of the Paper}
 
 The paper consists of two main parts.
 
 In the first part, we study relativistic RLM on flat space in light-front coordinates. The use of light-front coordinates is motivated by the belief that  they are better suited to bound state problems (see \cite{turgut2018attractive} for more references on this), moreover the Tamm-Dancoff  truncation in these coordinates is possibly  a good approximation to the original theory as the vacuum structure is not affected by the interactions \cite{wilson}. It also helps us establish some of the technical tools in a familiar situation before the manifold case is introduced. We start by renormalizing the so-called Principal Operator and  compute a lower bound for the ground state energy. The manipulations in doing so are only justified after we show that the formal expression we find for the resolvent corresponds to a well-defined Hamiltonian and the Principal Operator is a holomorphic family of type-A.
 
In the second part, we study the model on a compact manifold. We use some relatively recent heat kernel estimates on compact manifolds for the lower bound on the ground state energy. As the essential ideas are very similar to the light-front version, most of the computational details are omitted in the main text, nevertheless,   they can be found in the Appendix.

\section*{Description of the Model}

In this introductory chapter, we give a description of the model independent of the background. 
As introduced above, in RLM, we model the system as a two level system interacting with a bosonic scalar field. The $N$ particle is depicted as the up state $\chi_\uparrow$, whereas the $V$ particle as the down state $\chi_\downarrow$.
The Hamiltonian for RLM can be written as the sum of the free and the interaction terms:
\bb
H=H_0+\mu \frac{1 - \sigma_3}{2} +H_I
\label{eq: hamiltonianfull}
\ee
where $H_0$ is the free Hamiltonian for the bosonic field, $H_I$ is the interaction Hamiltonian and $\mu$ is the mass difference between $\chi_\uparrow$ and $\chi_\downarrow$. The only allowed interaction is,
\begin{equation}
 N+\theta\leftrightarrow V \, .    
 \end{equation}
which implies that this formal Hamiltonian acts on a direct sum  $\mathcal{F}^{(n+1)}_B(\mathcal{H})\otimes \chi_\uparrow\oplus{\mathcal{F}^{(n)}_B(\mathcal{H})}\otimes \chi_\downarrow $, where $\mathcal{F}^{(n)}_B(\mathcal{H})$ is the bosonic Fock space of $n$ particles.

The interaction Hamiltonian $H_I$ is given in terms of the truncated field operators $\phi^+ (x)$ and  $\phi^- (x)$ evaluated at $a$, where the fixed two-level system is located,

\bb
H_{I} =    \lambda  \left[ \sigma_+ \phi^{(-)}(a) + \sigma_-
    \phi^{(+)}(a) \right] 
\ee
where $\lambda$ is the coupling constant and $ \sigma_{\pm} = \frac{1}{2} (\sigma_1 \pm i\sigma_2) $.
This Hamiltonian commutes with the operator
\begin{equation}
   Q=\int dx dx^\perp :\phi^\dagger (x, x^\perp) \phi (x, x^\perp): -{1\over 2}(1+\sigma_3),
 \end{equation}
 which implies that either we have $n+1$ bosons and the down state or $n$ bosons and the up  state.

 \renewcommand{\thesection}{Part \Roman{section}:}

\setcounter{secnumdepth}{4}

\titleformat{\paragraph}
{\normalfont\footnotesize\bfseries}{\theparagraph}{1em}{}
\titlespacing*{\paragraph}
{0pt}{2.00ex plus 1ex minus .2ex}{1.5ex plus .2ex}

 \renewcommand{\thesubsection}{ \Roman{section}.\arabic{subsection}}
\section{Lee model in the Light-front Coordinates}

Lee model is one of the simplest field theory models where the renormalization can be done non-perturbatively. Despite the simplicity of the model, the exact spectrum of the model is difficult to obtain.  The resolvent formulation, which can be written down exactly for this model, first introduced by Rajeev \cite{rajeev1999bound}, enables us to study the spectrum of the model without the need of an explicit formula for the renormalized quantum Hamiltonian. To understand this model in more depth, we recall some basic ideas which were originally worked out in \cite{kaynak2009relativistic}. To avoid the repetition we deliberately work with the light-front formalism, some aspects of which were also covered in an appendix of \cite{kaynak2009relativistic}.  Moreover,  it is believed that the light-front coordinates provide a better approximation when truncated field operators are used \cite{wilson}.
Let us briefly discuss the oblique coordinate system we use, which was recently used and reviewed in \cite{turgut2018attractive}.
In the oblique light-front coordinate system $u= t+x$ is chosen to be the evolution parameter, so $u$ is the light-font "time". Otherwise we keep the coordinates $x,y$ as before, but rename $y$ as $x^\perp$ (the tranverse coordinate).
As a result, the line element, the metric tensor and its inverse are given by,
\begin{equation}
ds^2 = du^2 - 2 du dx - (dx^\perp)^2  \,,
\end{equation}
\begin{equation}
g_{\mu \nu} = \begin{pmatrix}
1 & -1 & 0  \\
-1 & 0 & 0 \\
0 & 0 & -1  \\

\end{pmatrix}\,, \quad g^{\mu \nu} = \begin{pmatrix}
0 & -1 & 0  \\
-1 & -1 & 0   \\
0 & 0 & -1   \\
 
\end{pmatrix} \,.
\end{equation}
The scalar product between the coordinate and the conjugate momenta is
\begin{equation}
p_\mu x^\mu = p_u u + p x + \pt  \xt \,
\end{equation}
where 
$p_u,p$ and $\pt$ are the light-front energy, the longitudinal and the transverse momenta, respectively. Note that we have down indices for momentum variables, as  it is more convenient to do so in the oblique formalism. We state the following results without  computations and for the details, refer to our recent work \cite{turgut2018attractive} where quantization of a scalar field is reviewed in detail. In the light-front (equal-time) formulation, the bosonic field operator has the expansion in terms of creation and annihilation operators,
\begin{equation}
\phi(x , \xt) = \lip \ipt \frac{1}{\sqrt{2 p}} \left [ \lap e^{- i p x - i \pt   \xt} + \ladp e^{ i p x + i \pt   \xt} \right] \,.
\end{equation}
Note that the longitudinal momentum $p$ only runs through positive values.  

Recall that,
\bb
H=H_0+\mu \frac{1 - \sigma_3}{2} +H_I
\ee
\noindent The free Hamiltonian for the bosonic field is:
\begin{equation}
H_0 = \lip \ipt \lop \ladp \lap \,, \\
\end{equation}
where $\lop = \frac{m^2 + p^2 + \pt^2}{2 p}$, whereas the interaction part is:
\bb
H_{I} =    \lambda  \left[ \sigma_+ \phi^{(-)}(0) + \sigma_-
    \phi^{(+)}(0) \right] \,.
\ee
 where, $\lambda$ is the coupling constant. Note that in this coordinate system the minimum value $m$ of $\omega$ corresponds to $p=m, p_\perp=0$, which is an advantage over the usual formalism.
 

The positive and the negative frequencies at zero light-front time  evaluated at the origin are:
\begin{align}
	\phi^{(+)}(0) &= \lip \ipt \frac{\lap}{\sqrt{2 p}} \,, \\
	\phi^{(-)}(0) &= \lip \ipt \frac{\ladp}{\sqrt{2 p}} \,.
\end{align}
and the bosonic $n$-particle  wave functions in the coordinate space are given by:
\begin{equation}
\ket{\psi}=\frac{1}{\sqrt{n!}} \int dx_1dx^\perp_1...dx_ndx^\perp_n \psi(x_1, x_1^\perp,...,x_n,x^\perp_n)\phi^{(-)}(x_1,x_1^\perp)...\phi^{(-)}(x_n,x_n^\perp) \ket{0}\nonumber
,\end{equation}
with $\psi$ being symmetric on all particle coordinates.
Moreover, we use a relativistically invariant norm; the Hilbert space norm for bosons in the momentum space  decomposition for $n$ particles is given by:
\begin{eqnarray}
   \langle \psi|\psi \rangle
&=&\int \frac{dp_1 dp_{1\perp}}{4\pi^2 }...\frac{dp_n dp_{n\perp}}{4\pi^2}\frac{|\psi(p_1,p_{1\perp},...,p_n,p_{n\perp})|^2}{2^np_1...p_n}
 .\end{eqnarray}
 The last expression is manifestly positive definite and the Fock space consists of finite norm wavefunctions.

 \subsection{Principal operator}\label{princ_op}
We introduce a brief presentation of the principal operator and refer to \cite{kaynak2009relativistic} for the details. The idea of the method is to calculate the resolvent by means of a formal identity introduced by Rajeev. If we write the Hamiltonian in $2\times 2$ decomposition according to up and down states\cite{rajeev1999bound} :
\bb
H - E = 
\begin{bmatrix}
H_0 - E & \lambda \phi^{(-)} (0) \\
\lambda \phi^{(+)} (0) & H_0 - E + \mu
\end{bmatrix} 
= 
\begin{bmatrix}
a & b^{\dagger} \\
b & d
\end{bmatrix}
\label{eq: H-E}
\ee
This formal Hamiltonian acts on a direct sum  $\mathcal{F}^{(n+1)}_B(\mathcal{H})\otimes \chi_\uparrow\oplus{\mathcal{F}^{(n)}_B(\mathcal{H})}\otimes \chi_\downarrow $, without mixing different sectors--due to the conservation law with charge $Q$. 
Thus each sector can be studied independently, consequently, the Hamiltonian is restricted to such sectors. 
The resolvent, as the formal inverse of $H-E$ is:
		\bb 
	R (E) =
	\begin{bmatrix}
		\alpha & \gamma \\
		\beta & \delta
	\label{eq: Rsimple}
	\end{bmatrix}
	\ee
		where:
	\begin{eqnarray}
	\nonumber
	\alpha &=& a^{-1} + a^{-1} b^\dagger ~ \Phi ^{-1} (E) ~ b a^{-1} \\
	\nonumber
	\beta &=& - \Phi^{-1} (E)~ b a^{-1} \\
	\gamma &=& -a^{-1} b^\dagger ~ \Phi^{-1} (E) \\
	\nonumber
	\delta &=& \Phi^{-1} (E) \\
	\nonumber
	\Phi &=& d - ba^{-1} b^\dagger
	\label{eq: abgd}
	\end{eqnarray}
and $\Phi$ is defined as the \textit{Principal Operator}. We have a formal expression for the resolvent which involves the 
inverse of the free bosonic Hamiltonian as well as $\Phi$. The free resolvent is well defined as long as $\operatorname{Im} (E) \neq 0$  or if $\operatorname{Im}(E)=0$, for  $E$   less than $(n+1)m$.
When searching for the bound states, one looks for the poles of the resolvent below the free bosonic spectrum. Hence in (\ref{eq: Rsimple}), these poles can only appear as zeros of $\Phi(E)$. This leads to a great simplification; the solutions of the eigenvalue equation $\Phi(E)\ket{\omega(E)} =0$  determine the corresponding $E$ values: if they are below the free spectrum one has bound states. The price one needs to pay for this simplification is that a linear eigenvalue problem is turned into a nonlinear one, since in general $\Phi(E)$ turns out to be a complicated function of $E$. We remark that in all  blocks of the resolvent, $\Phi^{-1}(E)$  operates always on the bosonic 
Fock space of $n$ particles $\mathcal{F}^{(n)}_B(\mathcal{H})$.

After normal-ordering the creation and annihilation operators in the operator  $\Phi(E)$, we face a divergent term; hence at this stage, we think of $\Phi(E)$ with a cut-off on the momentum variable $p$. To renormalize $\Phi(E)$, one can let $\mu$ depend on $p$ and choose it such that it cancels the divergence. We take one step further and impose the physical mass condition:

\begin{equation}
\Phi_R (E=\mu_p) \ket{0} = 0
\label{eq: masscond}
,\end{equation}

 where $\mu_p$ is interpreted as the physical binding energy for the composite, namely  $\chi_\downarrow$, which is the vacuum sector for the principal operator. This choice fixes the finite part to be $\mu_p$.
Now, one can write the renormalized $\Phi(E)$ as,
\begin{align} 
	\Phi_R(E) &=  \left.  (H_0 - E + \mu_p) \bigg[  1 +\lambda^2 \lip \ipt  \frac{1}{2p} \frac{1}{(\lop- \mu_p)(H_0-E+\lop)}\bigg] \right.  
	\label{eq: phi4}
 \\
 \nonumber
	& \qquad - \left. \lambda^2 \lip\ipt\liq\iqt \frac{1}{2 \sqrt{pq} }   \ladq \frac{1}{H_0 - E  + \lop + \loq} \lap  \right .
	\end{align}
	
For simplicity, \textit {we will drop R-subscript and  often will write} $\Phi(E)$. 
Above form makes $\Phi(E=\mu_p)\ket{0}=0$ condition manifest.
The resolvent formula (\ref{eq: Rsimple}), with  $\Phi(E)$ as in  (\ref{eq: phi4}), defines 
$\frac{1}{H-E}$ for  our interacting system. Now, we aim to show that it is possible to associate a well-defined quantum Hamiltonian to this operator family and all the information about the system is contained in the resolvent.

To accomplish these tasks,
 we next compute the flow of eigenvalues on the real axis of $E$, below $nm +\mu$. It can be easily computed that  $\langle\frac{\partial  \Phi(E)}{ \partial E} \rangle < 0 $. By the Feynman-Hellman formula \cite{kato2013perturbation} we get:
\begin{eqnarray}
\frac{\partial \omega_k}{ \partial E} = \bra{\omega_k} {\partial \Phi(E)\over \partial E} \ket{\omega_k} \implies 
\frac{\partial \omega_k}{ \partial E}&<&0
,\end{eqnarray}
where $\Phi(E)\ket{\omega_k}=\omega_k(E)\ket{\omega_k(E)}$, i.e. the $k$th isolated eigenvalue of $\Phi(E)$. Thus, the eigenvalues flow monotonically.
This means  $\omega_k(E)=0$ has a unique solution and these solutions correspond to the possible bound states. Here, we formally assume that the operator $\Phi(E)$ has a discrete set of eigenvalues. Because of the flow of eigenvalues, in any sector,  $\omega_0(E)=0$ i.e. the zero of the lowest eigenvalue of $\Phi(E)$, then gives us the ground state and this observation can be used to find a lower bound for the ground state energy for a fixed number of bosons. 

The use of Feynman-Hellman formula relies on the assumption that the eigenvalues are differentiable functions of the parameter $E$. Indeed we  aim for much more: $\omega(E)$'s are actually holomorphic functions of the {\it complex} parameter $E$. 


\subsection{Lower bound on the ground state}
\label{subsec: lowbound}

In this section we review the lower bound for the ground state energy \cite{kaynak2009relativistic}. Note that, if there exists an $E_*$ s.t. on the half plane $\mathcal{D}$ where $Re(E) < Re(E_*)$ the Principal Operator $\Phi$ becomes invertible with a bounded inverse, as a result, one can conclude that the resolvent does not have a pole in $\mathcal{D}$. For simplicity, assuming that $\Phi$ is self-adjoint on the real axis (something to be justified later), it is enough to show that the operator is invertible along the real axis below a certain value, since we are  looking for zeros along the real axis. Moreover, we restrict our attention to the region where $\operatorname{Re}(E)<nm+\mu_p$: recalling that $\chi_\downarrow$ with no boson is the vacuum sector, we expect the bound state energies below $nm + \mu_p$. A variational proof of this claim on a compact manifold is provided in Part II.
Analyzing the $\Phi_R(E)$ operator (details are provided in Appendix \ref{section: app1}) gives the lower bound on ground state energy for an $n$ particle sector as, 

\begin{equation}
\boxed{E_{gr} \geq m(n-1)+\mu_P -\frac{\lambda^2 \pi n}{2}.}
\end{equation}

\subsection{Spectral Projections}

Let us now digress briefly on the  use of spectral projections to calculate the bound state wave functions. {\it If the resulting renormalized formal expression indeed corresponds to the resolvent of a well-defined self-adjoint quantum Hamiltonian}, we can calculate the ground state wave function via the contour integral,
 \bb
 \mathbf{P}_{\Psi_0}=-{1\over 2\pi i} \oint_{E_{gr}} dE\, R(E)
 ,\ee
where we assume that a small contour is picked around the isolated ground state $E_{gr}$. Although we do not know in our model at light-cone that the ground state is unique and corresponds to an isolated eigenvalue, as the two level system is fixed, this seems to be a reasonable assumption. This  assumption is indeed correct when we deal with a non-relativistic version of this model on a compact manifold as shown in \cite{erman2014nondegeneracy} (otherwise the jump along the spectral cut should be considered, but this possibility will not be considered here).
If $\Phi(E)$ has a unique lowest eigenvector, the point at which this eigenvalue vanishes, gives us the desired ground state energy, moreover the above formula reveals that the 
ground state of the original model is then  also unique. 
To find an explicit formula, it is essential to know that the family $\Phi(E)$
is self-adjoint holomorphic of type-A in the sense of Kato, this guarantees  the holomorphicity of the  corresponding eigenvalues and eigenvectors as  functions of $E$.
If we write the wave functions in the two component form,
\bb
\ket{\Psi_0}=\begin{pmatrix}
\ket{\Psi_0^{(n+1)}}\\ \ket{\Psi_0^{n}}
\end{pmatrix}
\ee
we have 
\begin{eqnarray}
\nonumber\ket{\Psi_0^{(n+1)}} &=& \Big[ -{\partial \omega_0(E)\over \partial E}\Big|_{E_{gr}} \Big]^{-1/2} (H_0-E_{gr})^{-1}\phi^{(-)}(0) \ket{\omega_0(E_{gr})}\\
\nonumber\ket{\Psi_0^n} &=& \Big[ -{\partial \omega_0(E)\over \partial E}\Big|_{E_{gr}} \Big]^{-1/2} \ket{\omega_0(E_{gr})}
,\end{eqnarray}
justification of which requires holomorphicity of the eigenvalue $\omega_0(E)$ and its associated eigenvector in complex $E$, as emphasized. 
Here, in the coordinate basis we express the wave function for lowest eigenvector,
\bb
\ket{\omega_0(E_{gr})}=\int dx_1...dx_n \psi_0(x_1,...,x_n)\phi^{(-)} (x_1)...\phi^{(-)}(x_n)\ket{0}
.\ee
We use $x_1,...,x_n$ as general coordinates, in fact the above expression is valid if we interpret these coordinates as our light-front variables as well as assuming them coordinates on a compact manifold as we do in a later section. We remark that we should interpret the operator expression in the wave function, in the coordinate representation, as
\bb
(H_0-E_{gr})^{-1} \phi^{(-)}(0)=\int_0^\infty ds \int  dx \phi^{(-)}(x)k_s(x,0) e^{-s(H_0-E_{gr})}, 
\ee
where $k_s(x,\bar x)=\int [dpdp_\perp] e^{-s\omega(p,p_\perp) } e^{-ip(x-\bar x)-ip_\perp(x^\perp-\bar x^\perp)}$ corresponds to a  relativistic version of  the heat kernel for the light-front model (an analogous  expression in the manifold case  is linked to the heat kernel  by the subordination identity as to be seen). 
 The normalization is preserved by these formulae (can be checked by a tedious calculation), that is we can see that the expressions above lead to
$$
\bra{\Psi^n_0} \Psi^n_0\rangle +\bra{\Psi_0^{(n+1)}}\Psi_0^{(n+1)}\rangle=\Big[ -{\partial \omega_0(E)\over \partial E}\Big|_{E_{gr}} \Big]^{-1}\langle \omega_0(E_{gr})| \Big[ -{\partial \Phi(E)\over \partial E}\Big|_{E_{gr}}\Big]|\omega(E_{gr})\rangle=1
.$$

\subsection{Resolvent Defining a Hamiltonian}
In order to establish the fact that the formal expression   actually defines a Hamiltonian, we borrow some ideas from the theory of semi-groups. 
Let us recall the definition of  a pseudo-resolvent family,
\begin{definition}
Let $\Delta$ be a subset of the complex plane. A family $J(\lambda)$, $\lambda \in \Delta$, of bounded linear operators on $X$ ($X$ being the Banach Space) satisfying:
\bb
J(\lambda)-J(\mu) = (\lambda - \mu ) J(\lambda) J(\mu) 
\label{eq: reseq}
\ee
is called a pseudo-resolvent on $\Delta$ .

\begin{thm}
\cite{pazy2012semigroups}  Let $\Delta$ be an unbounded subset of $\mathbb{C}$ and let $R(E)$ be a pseudo-resolvent on $\Delta$. If there is a sequence $E_k \in \Delta$ such that $ |E_k | \rightarrow \infty $ as $ k \rightarrow \infty $ and
\bb
\lim_{k \rightarrow \infty} E_k R(E_k) x = -x ~ ~ for ~ all ~ x \in X 
\label{eq: decaycond}
\ee
then R(E) is the resolvent of a unique densely defined closed operator H.
\label{thm: decay}
\end{thm}
\end{definition}
The resolvent  we introduce does indeed satisfy the resolvent identity (as can be explicitly checked, we outline the main steps in the next part when we discuss the model on a manifold,  the verification of which is essentially algebraic). We  show that there exists an operator $H$, such that the resolvent $R(E)$ is the resolvent family of $H$, where $R(E)=\frac{1}{H-E}$ by means of the above decay conditions. Note that our initial (ill-defined) Hamiltonian is operating on $\mathcal{F}_B^{(n+1)}(\mathcal{H})\otimes \chi_\uparrow\oplus\mathcal{F}_B^{(n)}(\mathcal{H})\otimes \chi_\downarrow$. Therefore the resolvent is defined over this Hilbert space that we call $\mathcal{H}_{Q}$. The decay conditions of the above theorem are verified in Appendix \ref{decay-cond}.

 Thus, we conclude that $R(E)$ is the resolvent of a quantum Hamiltonian (which we cannot write down explicitly). 
 Let us emphasize that this already implies that $R(E)$ is the resolvent family of a self-adjoint operator if we can justify that 
 $R^\dagger (E)=R(\bar E)$ for complex values of $E$. This is formally true, but we must justify the formal operation of 
 $\Phi^\dagger(E)=\Phi(\bar E)$ carefully, since unlike the resolvent itself this is an unbounded operator. 
 In fact its inverse shows up in the resolvent formula, and we do not have an explicit expression for this inverse, in any case,  we must show that $\Phi(E)$ is a self-adjoint holomorphic family of type-A in the sense of Kato (for $\operatorname{Re}(E)<nm+\mu_P$) to justify many formal manipulations that we perform with $\Phi(E)$  (one consequence of which is  to verify the formal equality above). Thus we turn to this issue now.
 It is in some sense more abstract and technical but essential to justify the spectral projection formula alluded above as well.

\subsection{Holomorphic Structure and Self-Adjointness of the Principal Operator}
Here we introduce the concept of holomorphic family of type-A and show that indeed $\Phi(E)$ defines such a family. Moreover, there is a concept of self-adjointness for operator families defined over a complex domain as well, we show that this is actually true for our family.

\begin{definition}
A family T(E) $\in \mathcal{C} (X,Y)$ (closed linear operators from Banach spaces $X$ to $Y$) defined for $E$ in a domain $\Omega$ of the complex plane is said to be holomorphic of type-A if:
\begin{itemize}
\item $D (T(E)) = D$ is independent of $E$,
\item $T(E)u$ is holomorphic for $E \in \Omega$ for every $u \in D$. 
\end{itemize}
\label{def: hol}
\end{definition} 

\subsubsection{Finding the Common Domain}

To start we first find a common domain for the family $\Phi(E)$, most reasonable  choice seems to be $D(H_0)$. To justify $D(H_0)$ to be the common domain of $\Phi(E)$, we first express it as a product 
\begin{equation}
    \Phi(E)=(1+\tilde K(E)+\tilde U(E)) (H_0-E+\mu_p)
    \end{equation}
    with $\tilde K(E), \tilde U(E)$ being bounded operators (for complex $E$ in some domain). 
     From the analysis presented in Appendix \ref{common_domain_lf}, we see that both $  \tilde K (E) $ and $ \tilde U (E) $ are bounded. Additionally,  if Re(E) is chosen to be sufficiently small, then the norms of the operators $  \tilde K (E) $ and $ \tilde U (E) $ become less than $1/4$ and as a result the operator sum becomes invertible.

\subsubsection{Operator Family $\Phi(E)$ is Closed on its Common Domain}
We remind the reader the definition of  a closed operator:
\begin{definition}
An operator T is said to be closed if, for any sequence $x_k$ in its domain $D(T)$, $x_k\rightarrow x$ and $Tx_k \rightarrow y$ implies that $Tx = y$.
\end{definition}
We want to show that $\Phi(E)$ is closed in its domain $D(\Phi(E)) = D = D(H_0)$. Let us suppose that  we have a sequence $x_k\in D(H_0)$ that converges to $x$ as well as,
\bb
\Phi(E)  ~ x_k ~ \rightarrow ~ y
\ee 
When $\operatorname{Re}(E) \leq \operatorname{Re}(E_*)$ where $\operatorname{Re}(E_*)$ is sufficiently small,  such that $\Phi(E)$ becomes invertible  (notice that due to the  bound found above, there is a value for $E$ in the complex plane, below which the operator becomes invertible):
\begin{eqnarray}
\nonumber
& \ & [1 + \tilde{K}(E) - \tilde{U}(E)]~(H_0+\mu_p-E) ~ x_k ~ \rightarrow ~ y ~~ , ~~ x_k ~ \rightarrow ~ x \\
& \implies & (H_0 + \mu_p -E) x_k ~ \rightarrow ~ [1+\tilde{K}(E) - \tilde{U}(E)]^{-1} ~ y ~~,~~ x_k ~ \rightarrow ~ x 
\end{eqnarray}
Since $H_0$ is closed on its domain:
\begin{eqnarray}
\nonumber
& \ & (H_0 + \mu_p - E) x_k ~ \rightarrow ~ (H_0 + \mu_p -E) ~ x ~ = ~ [1 + \tilde{K}(E) - \tilde{U}(E)]^{-1} ~  y \\
& \implies & y ~ = ~ [1 + \tilde{K}(E) - \tilde{U}(E)]~(H_0 + \mu_p - E) ~ x
\end{eqnarray}
Subsequently, for $\operatorname{Re}(E) \leq \operatorname{Re}(E_*)$, $\Phi(E)$ is closed. For $\operatorname{Re}(E) > \operatorname{Re}(E_*)$, we rearrange according to \ref{eq: phidif} :
\begin{align}
&\Phi(E) - \Phi(E_*) = T(E,E_*) (E_* - E)
\end{align}
Following the same procedure as in the previous section, we find that the difference is bounded as long as $\operatorname{Re}(E)<nm+\mu_p$.
Next, $T(E,E_*)$ is  bounded and since every bounded operator on a Hilbert space is closable, we conclude that for any fixed  value  $E_*$:
\begin{eqnarray}
\nonumber
[\Phi(E)-\Phi(E_*)] ~ x_k &\rightarrow &  [\Phi(E)-\Phi(E_*)]x ~~ \\
\nonumber ~~\Phi(E_*) x_k &=& [1+ \tilde{K}(E_*) - \tilde{U}(E_*) ] ~ (H_0-E_*+\mu_p) x_k \\
\nonumber
& \rightarrow & [1+ \tilde{K}(E_*) - \tilde{U}(E_*) ] ~ (H_0-E_*+\mu_p) x 
\end{eqnarray}
We can now choose $E_*$ as above to make the operator $\Phi(E_*)$ invertible and divide the operator as a sum of  {\it an invertible part} and {\it a bounded part}, add them up to see that:
\begin{eqnarray}
y = [(1+ \tilde{K}(E_*) - \tilde{U}(E_*)) (H_0-E_*+\mu_p) + \Phi(E) - \Phi(E_*) ] x ~ \implies ~  \Phi(E) ~ x ~ = ~ y
\end{eqnarray}
Hence, we conclude that $\Phi(E)$ is closed on its domain $D(\Phi(E)) = D = D(H_0)$ (which is 
dense inside $\mathcal{F}_B^{(n)}(\mathcal{H})$).

\subsubsection{Holomorphicity of the Matrix Elements}
Since the principal operator $\Phi(E)$ is defined by an integral, it is not obvious that the result is holomorphic. Therefore, to establish the holomorphicity, we invoke the following theorem which is proved in \cite{everitt1997representation}:

\begin{thm}
Let $V \subseteq \mathbb{R}$ be a Lebesgue measurable set of positive or infinite measure, $\Omega$ be an open subset of $\mathbb{C}$ and $L^1(V)$ the Lebesgue integration space of complex valued functions on $V$. Define $\Theta(E) : \Omega \rightarrow \mathbb{C}$ by: \\
\begin{equation*}
\Theta(E) \defeq \int_V \phi(t,E)dt ~~,~~ E \in \Omega
\end{equation*}
where $\phi(t,E) : V \times \Omega \rightarrow \mathbb{C}$ satisfies:

\begin{itemize}
\item $\phi ~ (\cdot,E) \in L^1(V) ~~ , ~~ E \in \Omega $
\item $\phi ~ (t, \cdot) \in \mathbf{H}(\Omega) ~~ , ~~ t\in V$
\end{itemize}
where $\mathbf{H}(\Omega)$ denotes all  functions that are holomorphic on $\Omega$ and $t$ stands for all the parameters related to our product  measure (since there could be more than just one). If the mapping:
\begin{equation*}
E ~~ \rightarrow \int_V |\phi (t, E)| dt
\end{equation*}
is  bounded on  every compact subset of $\Omega$, then $\Theta(E)$ is holomorphic on $\Omega$.
\label{thm: holomorph}
\end{thm}

Since $\tilde K(E)$ can be defined by the spectral measure of $H_0$ and we have shown its boundedness for any choice of $E$ in the symmetric domain $\Omega$, its holomorphicity is easier to check (moreover a similar calculation for a compact manifold is to be presented in the next part). Therefore, we concentrate on the potential like part $\tilde U(E)$.

Let us write down explicitly the matrix elements of the operator $\tilde U(E)$ to see that it consists of integrable (over the parameter space) functions for all $E$ and holomorphic functions (for fixed parameters) in the variable $E$.
Note that once we isolate the piece $H_0-E+\mu_p$, which is clearly holomorphic in $E$, the preceeding operators  are actually bounded. The remaining matrix elements thus become,
\begin{align}
	    \langle\psi_1 |\tilde U(E)|\psi_2\rangle =  
	    \lambda^2 \int_{0}^{1} du 
	    \int_{0}^{\infty} s\,ds e^{-s(1-u)\mu_P}  \Big \langle \phi^{(+)}(g(su))\psi_1 \Big |e^{-s(H_0-E)} \Big|\phi^{(+)}(h(s))\psi_2 \Big \rangle.  
	\end{align}
Note that these are well defined expressions and for any fixed complex $E$, real part of which is below $nm+\mu_p$, they are finite.
To clarify our claims, we write the measure more explicitly  in this case (to exhibit the full parameter space),
\begin{eqnarray}
&\ &\!\!\!\!\! \Big \langle \phi^{(+)}(g(su))\psi_1 \Big|e^{-sH_0}\Big|\phi^{(+)}(h(s))\psi_2\Big\rangle= \int  {dp_2 dp_{2\perp}\over 4\pi^2 }...{dp_n dp_{n\perp}\over 4\pi^2} \int {dp dp_\perp \over 4\pi^2} {\overline{ \psi_2(p,p_\perp, p_2, p_{2\perp},...p_n,p_{n\perp})}\over 2^{n/2}\sqrt{pp_2...p_n}}{e^{-su\omega(p,p_\perp)}\over\sqrt{ 2p}}
\nonumber \\
&\ & ~~~~~~~~~~~~~~~~~~~~~~~~~~~~~~~~~~~~~~~~~~~~~~~~~~~~\times e^{-s\sum_{i=2}^n \omega(p_i, p_{i\perp})} \int {dqdq_\perp\over 4\pi^2}{\psi_1(q,q_\perp,p_2, p_{2\perp},...,p_n,p_{n\perp})\over 2^{n/2}\sqrt{qp_2...p_n}}{e^{-s\omega(q,q_\perp)}\over \sqrt{2q}}\nonumber
.\end{eqnarray}
If we choose any two measurable functions $\psi_1,\psi_2$, the above expression defines a measurable function over all the variables, since the full expression is absolutely integrable as shown above,  as a result,  we obtain a product (Lebesgue) measure over $s,u$ and all the momentum variables we have. 
Using the  theorem stated above, we establish that $\tilde U(E)$ defines a holomorphic family of type-A in the sense of Kato  for 
$\Omega=\{ E\in \mathbf{C}| \operatorname{Re}(E)<nm+\mu_P\}$. This in turn, combined with $\tilde K(E)$ claim, proves that $\Phi(E)$ is a holomorphic family of type-A. We need to establish that this family is self-adjoint (in the appropriate sense).

\subsubsection{Self-Adjointness of the Family $\Phi(E)$ }
\label{section: selfadj}

Note that, formally, $\Phi^\dagger (E) = \Phi(\bar{E})$, then at least, $D(\Phi(E)) \subset D(\Phi^\dagger (E))$. But to conclude self-adjointness, we need to show that they admit the same domain.

Our strategy will be the following: we  make use of the well-known Kato-Rellich Theorem \cite{reed1975ii} to show that $\Phi(E)$ is self-adjoint on some region on the real axis for $E$ chosen to be sufficiently small and then employ W\" ust's theorem \cite{wust1972holomorphic} (quoted below) to generalize it to the whole symmetric open domain of interest. 

\begin{thm}
Let $A : D(A) \rightarrow \mathcal{H}$ be a self-adjoint operator and $B: D(B) \rightarrow \mathcal{H}$ be symmetric. For $D(A) \subset D(B)$, if the following is satisfied:
\bb
||Bx|| \leq a||Ax|| + b||x|| ~~~, ~~~ \forall x \in \mathcal{H}
\ee
with $a < 1$, $b < \infty$; then $A+B : D(A) \rightarrow \mathcal{H}$ is self-adjoint.
\label{thm: kato}
\end{thm}

A proof of this well-known theorem is in the classical reference by Reed and Simon \cite{reed1975ii}.  Recall the form of the Principal Operator for $E$ :
\begin{eqnarray}
\nonumber
\Phi(E) &=& (1 + \tilde K(E) - \tilde U(E))(H_0 - E + \mu_p) \\
&=& \underbrace{(H_0 - E + \mu_p) + \tilde K(E) ( H_0 -E + \mu_p)}  \underbrace{-\tilde U (E) ( H_0 - E + \mu_p)} \\
\nonumber
&\ & ~~~~~~~~~~~~~~~~~~~~~~~ A ~~~~~~~~~~~~~~~~~~~~~~~~~~~~~~~~~~~~~B
\end{eqnarray}
If $A$ is invertible, we can write $x= A^{-1}y$ for some $y \in \mathcal{H}$ , so the inequality above becomes:
\bb
||BA^{-1} y|| \leq a||y|| + b||A^{-1}y||
\ee
We will work on the real axis where $E< E_*$, $E_*$ chosen to be sufficiently small such that $\tilde K(E)$ is strictly positive.  By the spectral theorem it is  self-adjoint (being a  continuous function of $H_0$) and well defined on the original domain $D(H_0)$. Hence $A$ is a self-adjoint operator on $D(H_0)$. It takes some work to justify that $B(E)$ is symmetric, since for continuous variables creation-annihilation operators are actually distributional valued. It is more natural to think of this expression via the following representation, 
\begin{equation}
    B(E)=\int_0^\infty  ds\, \phi^{(-)}( f(s))e^{-s(H_0-E)}\phi^{(+)}(f(s))
\end{equation}
where 
$$
  \phi^{(-)}(f(s))=  \lip \ipt \frac{\ladp}{\sqrt{2p}} e^{-s\omega(p, p_\perp)} 
  ,$$
similarly for the adjoint operator. The adjoint operation commutes with the integration since the integral is absolutely convergent. One can then show that this expression is indeed a symmetric operator for real values. (It is still a delicate matter to prove the equivalence of all these different  representation-of the same formal expression-a task which we leave to the reader).
Note that for $b=0$, if the following is true in some region:
\bb
||BA^{-1}|| < 1
\ee
then the conditions stated in Theorem \ref{thm: kato} are satisfied and $A+B$ is self-adjoint in that region. Rearranging:
\begin{eqnarray}
BA^{-1} &=& -\tilde U(E) (H_0 - E + \mu_p) [(1+ \tilde K(E)) (H_0 - E+ \mu_p)]^{-1}  -\tilde U(E) [1+\tilde K(E)]^{-1}
\end{eqnarray}
since   $\tilde K(E)$ is positive, 
\bb
||\tilde U(E) (1+ \tilde K(E))^{-1}|| \leq ||\tilde U(E)||
\ee
Recall that while searching for the bound on the ground state, we show that $||\tilde U(E)||<1$ if we choose $E<(n-1)m+\mu_p - C n\lambda^2$. In the same spirit, we can see that, from the estimate in  the complex case,  we can choose a sufficiently low value of $E$ (say less than or equal to $E_*$) on the real axis to make the above norm less than $1$.  Then,  $||BA^{-1}y|| \leq a$ where $a<1$ and  by the theorem statement,  $A+B= \Phi(E)$ is self-adjoint at least in some region where $E\leq E_*$. 

\begin{thm} {(W\"ust)}
Let $\Omega$ be a domain in the complex plane which is symmetric around the real axis and $\{ \Phi(E), ~ E \in \Omega \} $ be  a holomorphic family of type-A in 	$\mathcal{H}$ with dense domain $D_0$ such that $\Phi(\bar{E}) \subset \Phi^\dagger (E)$. 
Define $M$ by:
\begin{eqnarray}
M := \{ E ~|~ E\in U~, ~~ \Phi^\dagger (E) = \Phi(\bar{E}) \}
\end{eqnarray}
If $M$ is not empty, it extends to all of $\Omega$; i.e. $M \neq \emptyset  \implies M = \Omega$.
\label{thm: wust}
\end{thm}

The formal relation $\Phi^\dagger(E)=\Phi(\bar E)$ implies that domain inclusion alluded in the theorem holds.
As we have shown previously that at least in some region on the real line below a sufficiently small $E_*$, $\Phi(E)$ is self-adjoint. Thanks to the W\"ust's theorem, the equality (not only formally but in the real sense; meaning that domains are also equal) $\Phi^\dagger (E) = \Phi(\bar{E})$ extends to all $\{E\in \mathbb{C}|\operatorname{Re}(E)<nm+\mu_p\}$. Hence we conclude that $\Phi(E)$ is a  self-adjoint holomorphic family of type-A on the domain of interest.

\section{Lee model on 2D compact Riemannian manifolds}

In this second part of the paper, we analyze a  version of the Lee model where the two level system is fixed on a Riemannian manifold  interacting with an arbitrary number of bosons. Again,  we  employ the  nonperturbative renormalization method proposed by Rajeev \cite{rajeev1999bound}, where the resolvent is expressed in terms of the ``Principal Operator" $\Phi(E)$. Once  a finite expression for $\Phi(E)$ is found, the spectral information can be obtained from it. The zeros of the eigenvalues of $\Phi(E)$, as discussed in the previous part, correspond to the bound states  of the quantum Hamiltonian, if they are below the free spectrum. The compact version is technically much simpler, thanks to spectral results known in the case of compact manifolds. We plan to investigate uniqueness of the ground state only for this compact version in a forthcoming work.
To apply this technique to our model, we make use of an essential mathematical tool; the heat kernel. We  give a brief overview of this tool on Riemannian manifolds here not to interrupt the flow of the main work. Here we follow \cite{chavel1984eigenvalues}.

The heat equation on a Riemannian manifold $\mathcal{M}$ is given as:
\begin{equation}
  {\partial u\over \partial t}=-\nabla_g^2 u  
,\end{equation}
where $-\nabla_g^2$ is the Laplace-Beltrami operator on $\mathcal{M}$.
The keat kernel $K_t (x,y)$, defined on $(0,\infty) \times \mathcal{M} \times \mathcal{M}$, is a fundamental solution of the heat equation which is $C^2$ with respect to $x$ and $C^1$ with respect to $t$ satisfying:
\begin{eqnarray}
{\partial \over \partial t} K_t(x,y) = [-\nabla_g^2]_x K_t(x,y) ~~~~~;~~~~~ \lim_{t \rightarrow 0^+} K_t(\cdot,y) = \delta_{\cdot,y}
,\end{eqnarray}
as well as being positive and symmetric and it satisfies the semi-group property respectively:
\begin{eqnarray}
K_t(x,y) &=& K_t (y,x) \\
K_t(x,y) > 0 ~~ &;& ~~ x,y \in M ~~,~~ t \geq 0 \\
\int_M d_g z K_{t_1} (x,z) K_{t_2}(z,y) &=& K_{t_1+t_2} (x,y)
\end{eqnarray}
For $\mathcal{M}$ compact, there exists a {\it complete orthonormal basis} consisting of eigenfunctions $f_\sigma$ of the Laplace-Beltrami operator $-\nabla^2_g$ and the Sturm-Liouville decomposition of the heat kernel reads:
\begin{equation}
    K_t(x,y) = \sum _{\sigma} ^{\infty} e^{-\sigma t} f_\sigma(x) f_\sigma(y)
\end{equation}
where $\sigma$'s are the corresponding positive eigenvalues (counting multiplicities as well). It is essential that this {\it set of eigenvalues  is countable  with no accumulation point other than infinity}. 
The short time asymptotics for the diagonal heat kernel is given by:

\begin{equation}
K_t (x,x) \sim \frac{1}{(4\pi t)^{d/2}} \sum_{k=0} ^\infty a_k (x) t^{k} \quad t\to 0^+
\end{equation}
 where $d$ is the dimension of $\mathcal{M}$ and the smooth functions $a_k (x)$ (restricted to the diagonal) are given by explicit formulas in terms of local geometric invariants \cite{berline2003heat} with $a_0=1$. We can directly recognize the singular nature of the heat kernel near $0^+$. This is an important point to keep in mind throughout the work when giving estimates to some integrals and searching for the sources of possible divergences.

When estimating some expressions that we face, the following upper bound for the heat kernel on compact manifolds will be of great importance \cite{wang1997global}:
\begin{equation}
    K_t(x,x) \leq \frac{1}{V(\mathcal{M})} + Ct^{-d/2}
\label{eq: hkestimate}   
\end{equation} 
for all $t>0$ and $x \in \mathcal{M}$ where $d={\rm dim}(\mathcal{M})$, $V(\mathcal{M})$ is the volume of the manifold and $C$ is a positive constant which can be computed explicitly in terms of geometric invariants.
 We introduce the resolvent in terms of the Principle Operator $\Phi(E)$ as before. We explicitly construct $\Phi(E)$ and observe that the bound state solutions come from the poles of $\Phi(E)^{-1}$. Identifying  the divergence, we first put a cut-off to the allowed eigenvalues of the Laplacian and let the mass difference $\mu$ depend on $\Lambda$. Imposing the physical mass condition and solving for $\mu (\Lambda)$, we remove the divergence then take the limit $\Lambda \rightarrow \infty$ (as mentioned in the first part this process is not essential, we actually only need the resulting expression). \par
Once we establish a finite principal operator, we start searching for upper and lower bounds to the ground state energy. In this part thanks to the compactness, the variational method is employed to show that the ground state energy is indeed below the trivial guess $nm+\mu_p$ where $n$ is the number of bosons and $\mu_p$ is the physical binding energy. 
$E_*$ below which the Principal Operator is observed to be invertible serves as a lower bound to the ground state energy. 
As shown in the previous part, we  establish  that $R(E) = \frac{1}{H-E}$ is indeed the resolvent of a densely defined closed operator.
Subsequently, we study the holomorphicity of the Principal Operator. To show that $\Phi(E)$ is a self-adjoint holomorphic family of type-A in the sense of Kato, we fix the  common domain $D(H_0)$ as is done in the previous part and show that $\Phi(E)$ is closed on it.  We  conclude that $\Phi(E)$ is a self-adjoint holomorphic family of type-A following the same method as before. \par

\subsection{Hamiltonian and the Renormalized Resolvent}

For the relativistic Lee model on a 2+1 dimensional $compact$ Riemannian manifold $(\mathcal{M},g)$ , the formal Hamiltonian is (for more details see \cite{kaynak2009relativistic}):
\begin{equation}
   H= H_0 +\mu  {1-\sigma_3 \over 2} + H_{I} 
\end{equation}
Here, we have the free Hamiltonian,
\begin{equation}
H_0 = \sum_\sigma \omega_\sigma 
a^{\dagger}_\sigma a _\sigma 
\end{equation}
in a similar fashion, we have the interaction part,
\begin{equation}
H_{ I } = \lambda  \big[ \sigma_+ \phi^{(-)} (\bar{x} ) + \sigma_- \phi^{(+)}(\bar{x}) \big]
\label{eq: int}
\end{equation}
where $ \sigma_{\pm} = \frac{1}{2} (\sigma_1 \pm i\sigma_2) $ , $\omega_\sigma = \sqrt{\sigma + m^2} $ , $m$ the mass of the boson, $\bar{x}$ the location of the two level system. Compactness is not an essential restriction for the formalism presented below, but it simplifies the rigorous analysis we attempt in our work.

Since the manifold we are working on is compact, the Laplacian has a discrete spectrum and there is a family of orthonormal complete eigenfunctions $ f_\sigma (x) \in L^2 (M) $ which satisfy \cite{rosenberg1997laplacian} : 
\begin{eqnarray} 
\cr \int_{M} d_g x \; f_{\sigma}^{*}(x)\,
f_{\sigma'}(x)&=& \delta_{\sigma \sigma'}  \;, \cr \sum_{\sigma}
f_{\sigma}^{*}(x) f_\sigma(y)&=& \delta_g(x,y) \;
\label{eq: eigenbasis}
\end{eqnarray}
where $d_gx = \sqrt{{\rm det} [g_{ij}]} dx $ is the volume element and we introduce:

\begin{eqnarray} 
\cr \phi^{(-)} (x) & = & \sum_\sigma \frac{1}{\sqrt{2\omega_\sigma}} f^* _\sigma (x) a^\dagger _\sigma \,
\cr \phi^{(+)} (x) & = & \sum_\sigma \frac{1}{\sqrt{2\omega_\sigma}} f_\sigma (x) a _\sigma \
\end{eqnarray}
Since $f_\sigma (x)$ 's can be chosen to be real, the complex conjugate will not be important in the following calculations.
We use $n$-particle Hilbert space with an invariant norm as before,
\bb
\ket{\psi}= {1\over \sqrt{n!}}\int_{\mathcal M} d_gx_1...d_gx_n \psi(x_1,...,x_n) \phi^{(-)} (x_1)...\phi^{(-)}(x_n) \ket{0}
,\ee
where $\psi(x_1,...,x_n)$ is symmetric in all its entries.
The inner product can be written in a nicer form in the eigenfunction decomposition,
\bb
\bra{\psi} \psi\rangle= \sum_{\sigma_1,...,\sigma_n} {1\over    2^n } \frac{|\psi(\sigma_1,...,\sigma_n)|^2}{\omega_{\sigma_1}...\omega_{\sigma_n}}
\ee
or in coordinate space as
\bb
\bra{\psi} \psi\rangle= {1 \over 2^n} \int_\mathcal{M} dx_1...dx_n \overline{\psi(x_1,...,x_n)}[-\nabla^2_g+m^2]_{x_1}^{-1/2}...[-\nabla^2_g+m^2]^{-1/2}_{x_n}\psi(x_1,...,x_n)
.\ee
The coordinate version will not be used in this work, although for some other purposes, such as uniqueness of the ground  state coordinate space measure becomes important (this is because the ground state wave function is strictly positive in coordinate space representation). 

\subsection{Principal  Operator and Spectral Flow}

To construct a finite model,  we calculate the resolvent by an alternative method as is done for the light-front version,
and arrive  at the principal operator:
\begin{equation}
    \Phi(E) = [ H_0 -E + \mu ] - \sum_{\sigma , \tau} \frac{\lambda^2}{\sqrt{2\omega_\sigma}} f_\sigma a_\sigma \frac{1}{H_0 - E} \frac{1}{\sqrt{2\omega_{\tau}}} f_{\tau} a_{\tau} ^\dagger
\label{eq: phicutoff}
.\end{equation}
This is a formal expression and we now normal order this operator and  arrive at:
\begin{eqnarray}
\Phi_\Lambda (E) = \Big[H_0 - E + \mu(\Lambda) \Big] - \sum _{\sigma < \Lambda} \frac{\lambda^2}{2\omega_\sigma} |f_\sigma | ^2 \frac{1}{H_0 - E + \omega_\sigma}
\label{eq: philambda} 
- \sum_{\sigma , \tau < \Lambda} \lambda^2 f_\tau \frac{a^\dagger _\tau}{\sqrt{2\omega_\tau}} \frac{1}{H_0 - E + \omega_\sigma + \omega_\tau} \frac{a_\sigma}{\sqrt{2\omega_\sigma}} f_\sigma
\end{eqnarray}
Note that we introduce a cut-off anticipating a divergence; the second term in (\ref{eq: philambda}) diverges as $\Lambda \rightarrow \infty$. In order to make sense of all \textit{formal operations}, we should put a cut-off to the allowed eigenvalues of $-\nabla_g^2$. 
We  choose $\mu(\Lambda)$ such that we remove the divergence in (\ref{eq: philambda}). This still leaves out some ambiguity in the finite parts. But if we impose the condition that  we get a  zero when $E=\mu_p$ , where $\mu_p$ is the physical binding energy for the boson and the down state composite, which is  the vacuum sector for the principal operator, i.e. 
\begin{equation}
\Phi_R (E=\mu_p) \ket{0} = 0
\label{eq: cond}
,\end{equation}
this fixes the finite part to be $\mu_p$.
This is a renormalization condition typical of  such problems. As a result for  $\mu(\Lambda)$ we get :
\begin{equation}
    \mu (\Lambda) = \sum
_{\sigma < \Lambda} \frac{\lambda^2}{2\omega_\sigma} |f_\sigma| ^2 \frac{1}{(\omega_\sigma - \mu_p)} ~ + ~ \mu_p
\end{equation}
the principal operator becomes after the physical mass condition  imposed,
\begin{eqnarray}
\Phi(E) = (H_0-E+\mu_p) \Big[ 1 + \sum_{\sigma} \frac{\lambda^2}{2\omega_\sigma} \frac{1}{(H_0-E+\omega_\sigma)} \frac{f_\sigma ^2 (\bar{x})}{(\omega_\sigma -\mu_p)} \Big]
- \sum_{\sigma,\tau } \lambda^2 f_\sigma (\bar{x}) \frac{a^\dagger _\sigma}{\sqrt{2 \omega_\sigma}} \frac{1}{H_0 - E + \omega_\sigma + \omega_\tau} \frac{a_\tau}{\sqrt{2\omega_\tau}} f_\tau (\bar{x}) ~ .
\nonumber
\end{eqnarray}
Recall that the eigenvalues of $\Phi(E)$ carry essential information about the spectrum, 
\begin{equation}
    \Phi(E)\ket{\omega_k(E)} =\omega_k(E)\ket{\omega_k(E)},
\end{equation}
here we assume again that the eigenvalues are differentiable functions of $E$. As to be anticipated,  we prove later that they behave  better than that, they are actually holomorphic functions of $E$.
We now compute the flow of eigenvalues as we change $E$ {\it along the real axis} while staying below $nm + \mu_P$. This can be accomplished by means of Feynman-Hellman formula \cite{kato2013perturbation} (equation 3.18 page 391):
\begin{eqnarray}
{\partial \omega_k(E)\over \partial E} = \bra{\omega_k(E)} {\partial \Phi(E)\over \partial E} \ket{\omega_k(E)} &=& -1-{\lambda^2\over 2} \sum_{\sigma} \bra{\omega_k(E)} { f^2_\sigma(\bar x)\over \omega_\sigma(H_0-E+\omega_\sigma)^2}\ket{\omega_k(E)} \nonumber\\
&\ & -{\lambda^2\over 2} \underbrace{\sum_{\sigma\tau}\bra{\omega_k(E)}{a_\sigma^\dagger\over \sqrt{\omega_\sigma}} {f_\tau(\bar x) f_\sigma(\bar x)\over (H_0-E+\omega_\sigma+\omega_\tau)^2} {a_\tau\over \sqrt{\omega_\tau}}\ket{\omega_k(E)}} \nonumber\\
&\ & \ \ \ \  \int_0^\infty s ds \Big| \Big| \sum_\sigma e^{-s({1\over 2} H_0+\omega_\sigma -{1\over 2} E)}  
{f_\sigma(\bar x)a_\sigma\over \sqrt{\omega_\sigma}} \ket{\omega_k(E)} \Big| \Big|^2\nonumber\\
\implies 
{\partial \omega_k(E)\over \partial E}&<&0
\label{eq: lesszero}
\end{eqnarray}
\subsection{Upper Bounds on the Ground State}
\label{subsection: upper}

We want to show that there is an upper bound to the ground state energy by means of the variational principle. We choose a trial function:
\begin{equation}
\ket{\Omega_*} = \frac{1}{\sqrt{n!}} a^\dagger _0 ... a^\dagger _0 \ket{0}
\label{eq: trial}
\end{equation}
where we have $n$ creation operators with $\sigma = 0$. This is possible on a compact manifold since $ (- \nabla_g^2) \frac{1}{\sqrt{V(M)}} = 0$ is a constant solution \cite{rosenberg1997laplacian}, where $\frac{1}{\sqrt{V(M)}}$ is chosen for the sake of normalization $ \int |f|^2 dv ~ = ~ 1$   \\
The zero's of the principal operator give us bound state energies since they are the poles of the resolvent. Accordingly, if we  show that :
\begin{equation}
\omega_0 (E_*) \leqslant \bra{\Omega_*} \Phi_R (E_*) \ket{\Omega_*} <  0
,\end{equation}
where by the variational principle $\omega_0(E)$ refers to the smallest eigenvalue,
we can deduce, using ${\partial \omega_0(E)\over \partial E}<0$, that 
$$
E_{gr} < E_*
$$
Making a trivial guess, we set $E_*= nm + \mu_p$ , corresponding to the sector $Q=n+1$. 
\begin{eqnarray}
\nonumber
\bra{\Omega_*} \Phi_R (E_*) \ket{\Omega_*} &=& \bra{\Omega_*} (H_0 - nm)  \ket{\Omega_*} ~~~~~~~~~~~~~~~~~~\\
 &+& \bra{\Omega_*} (H_0 - nm ) \sum_{\sigma} \frac{\lambda^2}{2\omega_\sigma} \frac{|f_\sigma|^2 }{(H_0- nm - \mu_p +\omega_\sigma)} \frac{1}{(\omega_\sigma -\mu_p)} \ket{\Omega_*}  \nonumber \\
&-& \bra{\Omega_*} \sum_{\sigma,\tau} \lambda^2   \frac{f_\sigma a^\dagger _\sigma}{\sqrt{2 \omega_\sigma}} \frac{1}{(H_0 - nm - \mu_p + \omega_\sigma + \omega_\tau)} \frac{f_\tau a_\tau}{\sqrt{2\omega_\tau}} \ket{\Omega_*}
\label{eq: phivar}
\end{eqnarray}
The first part becomes zero as can be easily checked, the last ``potential" part gives
\begin{eqnarray}
- \lambda^2\bra{\Omega_*} \sum_{\sigma,\tau} \!\!\!\! & f_\sigma & \!\!\!\! \frac{a^\dagger _\sigma}{\sqrt{2 \omega_\sigma}} \frac{1}{(H_0 - nm - \mu_p + \omega_\sigma + \omega_\tau)} \frac{a_\tau}{\sqrt{2\omega_\tau}} f_\tau \ket{\Omega_*} \nonumber\\
&=& - \frac{\lambda^2}{n!} \bra{0} (a_0)^n  \sum_{\sigma ,\tau} f_\sigma f_\tau \frac{a_\sigma ^\dagger}{\sqrt{2\omega_\sigma}} \frac{1}{H_0-nm+\mu_p +\omega_\sigma +\omega_\tau} \frac{a_\tau}{\sqrt{2\omega_\tau}} (a_0 ^\dagger)^n \ket{0} \nonumber\\
\nonumber
&=& -n^2 \frac{\lambda^2}{n!} \bra{0} \underbrace{ a_0 ... a_0 }_{n-1} ~ \big( \frac{ |f_0| ^2}{m} \frac{1}{H_0 -nm + \mu_p +m +m} \big) ~ \underbrace{a_0 ^\dagger ... a_0 ^\dagger}_{n-1} \ket{0} \\
&=& - \frac{n  \lambda^2 }{m(m+\mu_p)} ~  |f_0|^2,
\end{eqnarray}
consequently the desired inequality is established.
Note that we have $\omega_0 (nm+\mu_P) < 0 $ and we know, $ \frac{\p \omega_0(E)}{\p E} < 0$ (Equation \ref{eq: lesszero}). Thus, we need to \textit{reduce} $E$ to get $\omega_0 (E_{gr}) = 0 $ which is the sought after result for the bound state energy. This implies the following inequality for the actual ground state energy:
\begin{equation}
\boxed{E_{gr} < nm + \mu_p}
\end{equation}

\subsection{Lower Bounds on the Ground State}
\label{subsection: lower}
Let us now think about a lower bound for the ground state energy. We write $\Phi(E)$ in a symmetrical form assuming real values of $E$ and the term,
\begin{equation}
   K(E)= \sum_{\sigma} \frac{\lambda^2}{2\omega_\sigma} \frac{1}{(H_0-E+\omega_\sigma)} \frac{f_\sigma ^2 (\bar{x})}{(\omega_\sigma -\mu_p)}
\end{equation}
is strictly positive, therefore it can be dropped,  which leads us to the following operator inequality provided that $E$ is real:
\begin{eqnarray}
\nonumber
\Phi_R &\geq& (H_0-E+\mu_p)^{1/2} \Big[ 1 -  \frac{\lambda^2}{2} \sum_{\sigma , \tau} \frac{f_\sigma a_\sigma ^\dagger }{  \sqrt{\omega_\sigma }}   \frac{1}{(H_0 - E +\mu_p +\omega_\sigma)^{1/2}}  \frac{1}{ (H_0 - E + \omega\sigma + \omega_\tau)}
 \frac{1}{( H_0 - E + \mu_p  + \omega_\tau)^{1/2}} \frac{f_\tau a_\tau}{\sqrt{\omega_\tau}} \Big]\\
 &\ & ~~~~~~~~~~~~~~~~~~~~~~~~~~~~~~~~~~~~~~~~~~~~~~~~~~~~~~~~~~~~~~~~~~~~~~~~~~~~~~~~~~~~~~~~~~~~~~~~~~\times (H_0 -E + \mu_p )^{1/2} 
\label{eq: aa}
\end{eqnarray}
Call the second term in the square brackets as $\mathcal{U}$. Then,
\bb
\Phi_R \geq (H_0 -E + \mu_p )^{1/2} ~ \big[ 
1 - \mathcal{U}(E) \big] ~ (H_0 -E + \mu_p )^{1/2} 
\ee
If $ ||\mathcal{U}(E)|| < 1$, the right hand side is  invertible and so is the Principal Operator. Therefore, if we can find $E_*$ below which $ ||\mathcal{U}(E)|| < 1$, we can deduce directly:
\bb
E_{gr} \geq E_*
.\ee
Let us  define $\chi = (n-1)m - E$. Noting that $H_0 \geq (n-1)m $, we can replace $H_0 - E $ 's by $\chi$ and bring $\mathcal{U}$ in to the operator inequality form we used in the previous part:
\begin{eqnarray}
|| \mathcal{U}(E) || \leq \frac{n \lambda^2}{2} ~ \Big[ \sum_{\sigma , \tau} \frac{|f_\sigma|^2 |f_\tau |^2}{\omega_\sigma \omega_\tau (\chi + \omega_\sigma) ( \chi + \omega_\sigma + \omega_\tau ) ^{2} (\chi +\omega_\tau) } \Big] ^{1/2}
\end{eqnarray}
where we have also omitted $\mu_p$'s for convenience. Using the crude inequality:
\bb
(\chi + \omega_\sigma + \omega_\tau) ^2 > (\chi + \omega_\sigma)(\chi + \omega_\tau)
\ee
we decouple $\sigma$ and $\tau$ to get:
\bb
|| \mathcal{U}(E) || \leq \frac{n \lambda^2}{2} ~  \sum_{\sigma} \frac{|f_\sigma|^2}{\omega_\sigma (\chi + \omega_\sigma)^2  }  
\ee
Using Feynman parametrization, exponentiation and subordination identity consecutively we get:
\begin{eqnarray}
\nonumber
|| \mathcal{U}(E) || \leq \frac{n\lambda^2}{2} \int_0 ^1 \xi d\xi \int _0 ^\infty \frac{s^3 ds}{2\sqrt{\pi}} e^{-s\xi \chi} \int_0 ^\infty u^{-3/2} e^{-s^2 / 4u} \underbrace{\sum_\sigma |f_\sigma|^2 e^ {-u \omega_\sigma ^2}} \\
\nonumber
~~~~~~~~~~~~~~~~~~~~~~~~~~~~~~~~~~~~~~~~~~~~~~~~~~~~~~~~~~~~~~~~~~~~~~~~~~~~
K_u (\bar{x} , \bar{x}) e^{-um^2}
\end{eqnarray}
Using the heat kernel estimate (\ref{eq: hkestimate}) we have:
\begin{eqnarray}
\nonumber
|| \mathcal{U}(E) || &\leq & \frac{n\lambda^2}{4\sqrt{\pi}} \int_0 ^1 \xi d\xi \int _0 ^\infty s^3 ds e^{-s\xi \chi} \int_0 ^\infty u^{-3/2} e^{-s^2 / 4u} \big( \frac{1}{V(\mathcal{M})} + \frac{C}{u} \big) e^{-mu^2} \\
\nonumber
&\leq & n \lambda^2 \Big\{ \int_0 ^1 \xi d\xi \frac{1}{V(\mathcal{M})} \frac{1}{(\chi \xi + m)^3}  ~ + ~ \int_0 ^1 \xi d\xi \frac{C (2m + \chi \xi)}{(m + \chi \xi)^2} \Big\} \\
&\leq & n \lambda^2 \Big\{ \frac{1}{V(\mathcal{M}) 2m (m+ \chi)^2} + \frac{C}{m+\chi} \Big\}
\nonumber \\
&\leq & n \lambda^2 \Big\{ \frac{1}{2m  V(\mathcal{M})\chi ^2} + \frac{C}{\chi} \Big\}
\end{eqnarray}
For $\chi > m$ , we can replace one of the $\chi$'s by $m$:
\bb
|| \mathcal{U}(E) || \leq n\lambda^2 \Big\{ \frac{1}{2m^2 V(\mathcal{M})} + C \Big\} \frac{1}{\chi}
\label{eq: ubound}
\ee
If we impose the condition:
\bb
\frac{n\lambda^2}{\chi} \Big\{ \frac{1}{2m^2 V(\mathcal{M})} + C \Big\} < 1 
\label{eq: condition}
\ee
$ || \mathcal{U} (E) || < 1 $ is guaranteed. Substituting $\chi = (n-1)m - E$  we get the lower bound for the ground state energy: 
\bb 
\boxed{(n-1)m - n\lambda^2 \Big( \frac{1}{2m^2 V(\mathcal{M})} + C \Big) ~ < ~ E_{gr}}
\label{eq: lowbound}
\ee
which was first presented in \cite{kaynak2009relativistic}.

\subsection{Resolvent Defining a Hamiltonian}
 
 As discussed in the previous part we need to check that $R(E)$ is a pseudo-resolvent, since we have the resolvent defined as a summation over the eigenmodes we will explicitly check it.
To show that $R(E) $ is a pseudo-resolvent, we need to check:
\bb
R(E_1) - R(E_2) \overset{?}{=} (E_1 - E_2) \Big( R(E_1) - R(E_2) \Big)
\label{eq: isres}
\ee
which is equivalent, according to two by two matrix form :
\begin{eqnarray}
\begin{bmatrix}
\nonumber
\alpha(E_1) - \alpha(E_2) & \gamma(E_1) - \gamma(E_2) \\
\beta(E_1) -\beta(E_2) & \delta(E_1) - \delta(E_2)
\end{bmatrix} 
= 
\begin{bmatrix}
\alpha(E_1)\alpha(E_2) + \gamma(E_1)\beta(E_2) & \alpha(E_1)\gamma(E_2) + \gamma(E_1)\delta(E_2) \\
\beta(E_1)\alpha(E_2) + \delta(E_1)\beta(E_2) & \beta(E_1)\gamma(E_2) + \delta(E_1) \delta(E_2)
\end{bmatrix}
\end{eqnarray}
(refer to (\ref{eq: abgd}) for the definitions of the terms). We remark that all operators are bounded here hence there are no issues about domains. In Appendix \ref{resolvent-manifold}, we show in detail that $R(E)$ indeed defines a resolvent operator.

\subsection{Holomorphic structure of the Principal Operator}
\label{chapter: hol}

It is well-known that to obtain a spectral decomposition of a family of operators in which eigenvalues and the corresponding projections are holomorphic functions of the parameter, we need the notion of a self-adjoint holomorphic family of type-A in the sense of Kato. This in turn justifies the fact that our resolvent formula defines a self-adjoint quantum Hamiltonian as well as putting our estimates on a firmer ground.  \\

Similarly as in Part I, we can   establish the following claim:
The family $\Phi(E)$,  defined for $\operatorname{Re}(E)<nm+\mu_p$, on a  symmetric domain of the complex plane is  holomorphic of type-A, that is
\begin{itemize}
\item $D (\Phi(E)) = D(H_0)$, independent of $E$,
\item $\Phi(E)$ is closed on this common domain,
\item $\Phi(E)u$ is holomorphic for $E \in D(H_0)$ for every $E$ in the open symmetric domain. 
\end{itemize}
Since the computation is very similar to Part {\bf I}, here we only outline the calculation steps.
We start by showing that the family can be given a common dense domain for $\operatorname{Re}(E)<nm+\mu_p$ on which it is closed. To establish self-adjointness of the family $\Phi(E)$, we rely on the W\"ust's theorem that it is enough to establish the self-adjointness condition even at a single point. This in turn is true due to Kato-Rellich theorem on self-adjointness when $E$ is sufficiently small on the real axis \cite{reed1978iv}. This part of the proof is essentially identical to the light-front case, since the proof given there is formal, it does not require the explicit forms of the operators.

\subsubsection{Holomorphicity of the Matrix Elements}

We now want to show that the family $\Phi(E)$ satisfies the second criteria in the Definition \ref{def: hol} of Part {\bf I}. Note that operator family $\Phi(E)$ is not given by an explicit formula it is an integral of a parameter dependent operator. To understand its holomorphic structure, by definition, it is essential to analyze the matrix elements. For this we can employ   Theorem \ref{thm: holomorph} (of Part {\bf I}) from  \cite{everitt1997representation} as used  in the previous part.

To make contact with this theorem, we note that
in our case, $\Theta(E) = \bra{\lambda} \Phi(E) \ket{\Psi}$ where $\ket{\lambda} \in  \mathcal{F}^{(n)} $ and $\ket{\Psi} \in D(H_0)$. For a family of  unbounded operators,  operator acts on the domain then we can take an inner product of the resulting vector with any vector in the Hilbert space (here it is essential that the family has a common domain for any value of the complex parameter $E$).
Recall that the Principal Operator reads as:
\begin{eqnarray}
\Phi (E) &=& \Big[ 1 + \sum_\sigma \frac{\lambda^2}{2\omega_\sigma} \frac{|f_\sigma|^2}{(H_0 - E + \omega_\sigma)} \frac{1}{(\omega_\sigma - \mu_p)} \\
\nonumber
& - & \sum_{\sigma,\tau} \lambda^2 f_\sigma \frac{a_\sigma^\dagger}{\sqrt{2\omega_\sigma}} \frac{1}{H_0 - E + \omega_\sigma + \omega_\tau} \frac{1}{H_0 - E + \omega_\tau + \mu_p} \frac{a_\tau}{\sqrt{2\omega_\tau}} f_\tau \Big] (H_0 - E + \mu_p)
\end{eqnarray}
$H_0-E+\mu_p$ is already  obviously holomorphic on the entire complex plane. We will again call the second term in square brackets as $\mathcal{K}(E)$ and the third as $\mathcal{U}(E)$. The main thing is to prove that these bounded operators are actually holomorphic in the desired open domain of the complex plane. Note that the operator $H_0+\mu_p-E$ is invertible for our choice of $E$, so its range is the full Hilbert space.
Therefore we show that for any choice of $\ket{\lambda}, \ket{\psi}$ the matrix elements
$\bra{\lambda} [1+\mathcal{K}(E)-\mathcal{U}(E)]\ket{\psi}$, apart from $1$,  when considered as an integral representation, 
becomes a sum of two pieces, each of which is defined over the same domain of the complex plane. They carry different measures, but 
they can be put into the form,
\begin{eqnarray}
&\ & \tilde \Theta(E)=\int_{V} \phi(E, t) d\mu(t)\quad {\rm with} \\
&\ & \phi(E, \cdot) {\rm \quad holomorphic\ for\ almost\ all\ } t \in V\\
&\ & \phi(\cdot, t) {\rm \quad\ measurable\ and\  integrable\ for\ any\  }  \operatorname{Re}(E)<nm+\mu_P\\
&\ & E\mapsto \int_V |\phi(E,t)|d\mu(t) \quad {\rm is\ a \ bounded\ map\ on\ any\ compact\ subset\ of\ } \operatorname{Re}(E)<nm+\mu_P
.\end{eqnarray}
Theorem \ref{thm: holomorph} then implies that the sum is a holomorphic function of $E$.

Let us verify these for $\mathcal{K}(E)$ first, using our previous estimates, we readily find that,
\begin{eqnarray}
\nonumber
\bra{\lambda} \mathcal{K}(E) \ket{\psi} &=&
\mathcal{C} \int \limits_{0 \leq u_2+u_3 \leq 1} du_2 du_3 ~  \int s^3 ds \int d \xi \frac{e^{-\frac{s^2}{4\xi}-m^2\xi}}{\xi^{3/2}}K_\xi(\bar x,\bar x) e^{s\mu_p u_2} \bra{\lambda} e^{-su_3(H_0-E)} \ket{\psi}\\
&=&  \int \limits_{0 \leq u_2+u_3 \leq 1} du_2 du_3 ~  \int ds \int d\xi~~ \phi(s,u_2,u_3,\xi, E)
\label{eq: lkpsi}
\end{eqnarray} 
where we use the heat kernel and subordination identity for the $f_\sigma(\bar x) e^{-s\omega_\sigma}$ to get the heat kernel, after this  we    collect everything into $\phi$ except the integral measures. To show integrability, we employ the well-known result,
let $|\phi| \leq g$, $\phi$ being a measurable function. If $g$ is integrable, so is $\phi$.
Taking the absolute value of $\phi(s,u_2,u_3,\xi,E)$, we get:
\begin{eqnarray}
\nonumber
\int du_2 du_3   \int ds \int d\xi~~ | \phi | &=&  \mathcal{C} \int du_2 du_3  \int s^3 ds \int d\xi\frac{e^{-\frac{s^2}{4\xi}-m^2\xi}}{\xi^{3/2}} e^{s\mu_p u_2} K_\xi (\bar{x},\bar{x})  ~|\bra{\lambda} e^{-su_3 (H_0 - E)} \ket{\psi}| \\
\nonumber
&\leq & \mathcal{C} ~ |\braket{\lambda} \psi |\int du_2 du_3  \int s^3 ds \int d\xi ~ K_\xi (\bar{x},\bar{x}) 
e^{s \mu_p u_2} \frac{e^{-\frac{s^2}{4\xi}-m^2\xi}}{\xi^{3/2}} e^{s\mu_p u_2}   e^{-su_3 (nm - \operatorname{Re}(E))} \\
&=& \int du_2 du_3   \int ds \int d\xi ~ g(s,u_2,u_3,\xi, E)
\label{eq: phimeasure}
\end{eqnarray}
where we have defined $|\phi|< g$ by estimating $H_0$ as $nm$ as usual and $E$ is taken to have a fixed value below $nm+\mu_p$. $\phi$ consists of well-defined continuous functions hence measurable. Note that the integral (\ref{eq: phimeasure}) is the same as (\ref{eq: Kforint}) up to some constants. Thus, we can estimate it following the same steps and show that it is bounded. For the integral (\ref{eq: Kforint}) is already shown to be finite in (\ref{subsection: decay}), so is (\ref{eq: phimeasure}); consequently, we can conclude that $\phi(\cdot,E)$ is indeed in $L^1$. Incidentally the true parameter space depends on the wave functions and the heat kernel coming from the exponential  of $H_0$, therefore we have a multiple integral over the compact manifold weighted by heat kernels. In fact, it is easier to directly establish holomorphicity of $e^{-s(H_0-E)}$ by this argument and not to think of these integrals  as part of the measure. For clarity this is the approach we take. 

Note that for the parameters $u_2,u_3,s,\xi$ fixed, $\phi$ is simply an entire function of $E$ where the only factor depending on $E$ is $e^{-s(H_0-E)}$. Thus, the holomorphicity of $\phi(t,\cdot)$ is straightforward. 
The extra condition stated in the above Theorem  does not require more work since we have already shown the integrability through $(\int du_2 du_3 ds d\xi | \phi |)$ being bounded above by $(\int du_2 du_3 ds d\xi ~ g)$, for $\operatorname{Re}(E) < nm+ \mu_p$. The explicit bound can be found in (\ref{eq: Kbound2}).

As the explicit construction  done above for $\mathcal{K}(E)$, for $\mathcal{U} (E)$ we  make use of some inequalities based on the bounds we found before. 
Now let us note that
\begin{eqnarray}
\bra{\psi_1}\mathcal{U}(E)\ket{\psi_2} = \lambda^2 \int_0 ^\infty s ds \int _0 ^1 du  e^{s[E-\mu_p(1- u)]}\bra{\phi^{(+)}(g(su))\psi_1} e^{-sH_0}  \ket{\phi^{(+)}(f(s))\psi_2}
\end{eqnarray}
The inner product is a well-defined measurable function of $s,u$ for fixed $\psi_1, \psi_2$, and moreover  it is bounded thanks to the norm inequality that  $|\bra{\psi_1} \mathcal{U}(E) \ket{\psi_2}| ~ \leq ~ ||\mathcal{U}(E)|| ||\psi_1||~|| \psi_2||$. The explicit form of the functions are cumbersome but it can be found (indeed we need the explicit expression  if we aim to show  compactness of these operators, this is a required step for uniqueness of the ground state, which we hope to discuss in a later publication). Overall   boundedness can be established  as before, for the sake of brevity we do not present these expressions.
(It is most natural to use the coordinate  representation of the inner product for $\mathcal{U}(E)$, then we express these in terms of product measures over $\mathcal{M}$). Holomorphicity in $E$ is again straightforward since the only function containing $E$ is an entire one, $e^{sE}$.
Since the boundedness of $\mathcal{U}(E) $ is   satisfied and integrability condition follows  in a similar way as for $\mathcal{K}(E)$, 
 we have shown that $\Phi(E)$ is indeed a holomorphic family of type-A on the domain of $\{E\in \mathbf{C} | \operatorname{Re}(E)<nm+\mu_p\}$
(with the common operator domain $D(H_0)$ over this set).
This is essential to establish the spectral projections via a contour integral  as emphasized  previously in Part {\bf I}.

\subsubsection{Self-Adjointness of $\Phi(E)$ }

Note that, formally, $\Phi^\dagger (E) = \Phi(\bar{E})$, hence at least, $D(\Phi(E)) \subset D(\Phi^\dagger (E))$. But to conclude self-adjointness, we need  to show that they admit the same domain. As it is done in the first part, 
we  make use of the well-known Kato-Rellich Theorem \cite{reed1975ii} to show that $\Phi(E)$ is self-adjoint on some region along the real axis for which $E$ is chosen to be sufficiently negative and then employ W\"ust's theorem \cite{wust1972holomorphic} to generalize it to the whole region of concern.

Note that the argument used in the preceding part about the light-front version is purely formal, therefore it carries over to the manifold case exactly. We identify $A$ and $B$ parts in the 
 Principal Operator as before.
We  work on the real axis,  and we show that (for some choices of $E$),
\bb
 ||BA^{-1}||= ||\mathcal{U}(E) (1+ \mathcal{K}(E))^{-1}|| \leq ||\mathcal{U}(E)||< 1
\ee
then the conditions stated in Theorem \ref{thm: kato} are satisfied. Note that, by the spectral theorem, $A(E)$ is a self-adjoint operator for real values of the parameter $E$ belonging to the symmetric region $\Omega$, defined on a domain $D(H_0)$.  Moreover, $B(E)$ is a symmetric operator for real values of $E$ (in the compact manifold case this is easier to see since creation and annihilation operators in energy representation are ordinary unbounded operators, many of the formal properties can be  justified).
Exactly the same arguments as before  shows that $\Phi(E)$ is self-adjoint along the real axis where $E<E_*$. 
 Now,  thanks to the W\"ust's theorem, $M=\Omega$, that is the equality $\Phi^\dagger (E) = \Phi(\bar{E})$ (not only formally but also the equality in the  sense of domains)   extends to all $\{E\in \mathbb{C}|\operatorname{Re}(E)<nm+\mu_p\}$. Therefore, we conclude that $\Phi(E)$ is a  self-adjoint holomorphic family of type-A on the domain of interest.

\section*{ Conclusion}
The relativistic Lee model is reanalyzed in the 2+1 dimensional oblique light-front coordinates in more detail. The resolvent formulation, developed by Rajeev, enables us to study the spectrum of this model, in particular the ground state energy can be estimated from below and above by analyzing the principle operator. We show tha the resolvent obtained by a formal process indeed corresponds to the resolvent of an operator. To establish self-adjointness and obtain spectral projections, we show that the principal operator as a family dependent on a complex parameter $E$ (in some symmetric domain) is  a self-adjoint holomorphic family of type-A in the sense of Kato. In the second part of the paper, these results are extended to the  model defined over a compact manifold by means of the heat kernel techniques.
In the near future, we plan to prove uniqueness (or the non-degeneracy) of the ground state for the compact case.

 \section*{Acknowledgements}
 We would like to note that the extension of the first part  to the compact manifold case is based on M. Unel's master thesis which was  completed while she was a graduate student at  Bogazici University,  Department of Physics. M. Unel acknowledges support from the Villum Foundation via the QMATH Center of Excellence.
 The authors  would like to thank B.T.Kaynak and F. Erman for discussions. The basic ideas of this project was conceived while O. T. Turgut was visiting J. Hoppe at KTH, Stockholm, he is grateful to J. Hoppe for these invitations and the scientific support he has received during all these years. We would like to thank M. Znojil, A. Michelangeli and G. Dell'Antonio  for their kind interest  to our work, moreover O.T. Turgut would like to thank them for their continual support over the years. Last but not least, O. T. Turgut would like to thank M. Deserno for arranging a long term visit to Carnegie-Mellon Physics Department which brought this work to a completion. 
 
\begin{appendices}
\section{Calculation Details of \ref{subsec: lowbound}}
\label{section: app1}
For the notational convenience in the following calculations, we rename parts of $\Phi$ as follows:

\begin{align} 
	\Phi_R(E) &= \underbrace{ \left.  (H_0 - E + \mu_p) \bigg[  1 +\lambda^2 \lip \ipt  \frac{1}{2p} \frac{1}{(\lop- \mu_p)(H_0-E+\lop)}\bigg] \right.} 
	\label{eq: phi4}
 \\
 \nonumber 
 & ~~~~~~~~~~~~~~~~~~~~~~~~~~~~~~~~~~~~~~~~~~~~~~~~~~~~ K(E)
 \\
 \nonumber
	& \underbrace{\qquad - \left. \lambda^2 \lip\ipt\liq\iqt \frac{1}{2 \sqrt{pq} }   \ladq \frac{1}{H_0 - E  + \lop + \loq} \lap  \right .}
\\
\nonumber 
 & ~~~~~~~~~~~~~~~~~~~~~~~~~~~~~~~~~~~~~~~~~~~~~~~~~~~~~~~~~~~ U(E)
	\end{align}

The full kinetic part can be rewritten as $K(E) = (H_0-E+\mu_p) + K_1(E)$, where $(H_o-E + \mu_p)$ is the free part and $K_1(E)$ is given by,

\bb
 K_1(E)=\lambda^2 \lip \ipt  \frac{1}{2p}\Big( \frac{1}{\lop- \mu_p}    
	 -   \frac{1}{H_0 - E  + \lop}\Big)
.\ee
To establish a lower bound on this term for real values of $E$, we first perform the $ \pt$ integral,
\begin{equation}
K_1(E)= \lambda^2 \lip \bigg(\frac{-1}{\sqrt{2p(H_0-E)+m^2+p^2}} + \frac{1}{\sqrt{p^2+m^2-2p\mu_P}} \bigg)
\end{equation}

Let us collect the  terms under  a common denominator and multiply the top and the bottom by \\ $\sqrt{2p(H_0-E)+m^2+p^2}+\sqrt{p^2+m^2-2p\mu_P}$ to find

\begin{equation}
\lip \bigg(-\frac{1}{\sqrt{2p(H_0-E)+m^2+p^2}} + \frac{1}{\sqrt{p^2+m^2-2p\mu_P}} \bigg) \times \frac{\sqrt{2p(H_0-E)+m^2+p^2}+\sqrt{p^2+m^2-2p\mu_P}}{\sqrt{2p(H_0-E)+m^2+p^2}+\sqrt{p^2+m^2-2p\mu_P}}\nonumber
\end{equation}
\begin{equation}
\geq \lip (H_0-E+\mu) \frac{p}{(2p(H_0-E)+m^2+p^2)\sqrt{p^2+m^2-2p\mu_P}}
\end{equation}
where in the last inequality we replaced the smaller term $p^2+m^2-2p\mu_P$ by the bigger one with $H_0-E$ to get a lower bound.
Let us separate the multiplicative factor $(H_0-E+\mu_P)$ for the time being.
For the remaining part
we use  Feynman  parametrization (removing the numerical  factor) to get:
\begin{align}
& \lip	\int_{0	}^{1} \frac{(1-u)^{-1/2}p }{(2up(H_0-E)+p^2+m^2-2p\mu_P(1-u))^{3/2}} du \geq
	 \lip \int_{0	}^{1} \frac{(1-u)^{-1/2}p }{(2up(H_0-E+\mu_P)+p^2+m^2)^{3/2}} du\nonumber \\
	&= \lip \int_{0	}^{1} \frac{(1-u)^{-1/2}p }{((p+\underbrace{(H_0-E+\mu_P)}_\text{a}u)^2\underbrace{-(H_0-E+\mu_P)^2u^2+m^2}_\text{b})^{3/2}} du
\end{align}
where we drop $2p\mu_P$ term in the last inequality to get a lower bound again.
As a result we write the above expression as,
\begin{equation}
 \lip \int_{0	}^{1} \frac{(1-u)^{-1/2}p}{\underbrace{((p+au)^2}_\text{x}+b)^{3/2}} = \int_{0}^{1}  (1-u)^{-1/2}   \int_{(H_0-E+\mu_P)u}^\infty \frac{d x}{2 \pi} \, \frac{x-(H_0-E+\mu_P)u}{(x^2+b)^{3/2}}
,\end{equation}
performing the $p$ and  $u$ integrations and multiplying the result with the left out $(H_0-E+\mu_P)$ term again, we arrive at (collecting all the  numerical/constant factors as $C_0$):

\begin{equation}
K_1(E) \geq C_0 \ln \bigg[\frac{H_0-E+\mu_P+m}{m}\bigg]
\end{equation}
where $C_0$ is a computable constant. Finally, we obtain the following lower bound on $K_1(E)$,

 \begin{equation}
K_1(E) \geq C_0 \ln \bigg[\frac{H_0-E+\mu_P+m}{m}\bigg]
\end{equation}
where $C_0$ is a computable constant.
Consequently the full kinetic part, $K(E)=(H_0-E+\mu_P)+K_1(E)$, satisfies the lower bound
\begin{equation}
    K(E)\geq (H_0-E+\mu_P)+ C_0\ln \bigg[\frac{H_0-E+\mu_P+m}{m}\bigg]
\end{equation}

For the principal operator to be invertible (for real values of $E$), it is sufficient to satisfy the condition 
 $$
 ||\tilde U(E)||=||K(E)^{-1/2}U(E)K(E)^{-1/2}||<1.
 $$ 
To estimate $||\tilde U(E)||$, we need the following operator inequality,
 \begin{equation}
    \Big|\Big| \int dP dQ F(P,Q) a^\dagger(P) a(Q)\Big|\Big|\leq n \Big[ \int dP dQ |F(P,Q)|^2\Big]^{1/2}
    \label{eq: opin}
 \end{equation}
Once we estimate all $H_0$ by their lower bounds $(n-1)m$ and employ the operator inequality \ref{eq: opin}, we get,

\begin{align}
\!\!\!\!\!\!\!\!||\tilde U(E)|| &  \leq \lambda^2 n\bigg[\int_0^\infty {dpdq\over 4\pi^2}\int {dp_\perp dq_\perp\over 4\pi^2} \frac{1}{4pq} \frac{1}{(\Delta+\lop+ \loq)^2(\Delta+\lop )(\Delta+ \loq)}  \bigg]^{1/2} \\
&\leq  \frac{\lambda^2}{2}  \frac{ \pi n}{m(n-1)+\mu_P -E}\label{eq: U bound}.
\end{align}
We then impose the condition,
 \begin{equation}
 	   \frac{\lambda^2}{2}   \frac{ \pi n}{m(n-1)+\mu_P -E} <1
 	  ,\end{equation}
which gives  the lower bound on ground state energy for $n$ particle sector as the $\Phi(E)$ becomes invertible by a Neumann series (for real $E$).


\section{Verifying the decay conditions in light front coordinates}\label{decay-cond}

In order to show that $R(E)$ satisfies Theorem 1, we pick a sequence $\lambda_k$ on the negative real axis for every $k$, $\lambda_k <0< E_{gr}$. Since $\lambda_k = -|\lambda_k|$, the condition (\ref{eq: decaycond}) becomes:
\bb
\lim_{k \rightarrow \infty} \Big| \Big| [|\lambda_k| R(-|\lambda_k|) -1] x \Big| \Big| _{\mathcal{H}_Q}  = 0
\ee
And using the triangle inequality repeatedly we get:
\begin{eqnarray}
&\ & \Big\| \Big[ |\lambda_k| R(-|\lambda_k|) -1 \Big] \begin{pmatrix}
\ket{f^{n+1}} \\
\ket{f^n}
\end{pmatrix}
\Big\|_{\mathcal{H}_Q} 
\leq || ~ \big( |\lambda_k| \alpha(-|\lambda_k|) -1 \big)\ket{f^{n+1}} ~ || + ||~ |\lambda_k| ~ \gamma(-|\lambda_k|) \ket{f^n}|| \\ 
~&\ &~~~~~~~~~~~~~~~~~~~~~~~~~~~~~~~~~~~~~~~~~~~~~~~~~~~~~~~~~~ + ~ || ~|\lambda_k| \beta(-|\lambda_k|) ~ \ket{f^{n+1}} || + ||~ \Big( |\lambda_k| \delta(-|\lambda_k|)-1 \Big) \ket{f^n} ||  \label{eq: line2}\nonumber
\end{eqnarray}

Thus it is sufficient to show that as $k \rightarrow \infty$ all the terms on the right hand side of the inequality go to zero.
To establish this we need a more detailed analysis of the behaviour of the operator $\Phi(E)$ for large negative values of $E$.
	We know that $(H_0-E)^{-1}$ is the resolvent family of the free bosonic part, thus it satisfies,
	$$
	\Big[|\lambda_k| (H_0+|\lambda_k|)^{-1} -1\Big]\ket{f^{n}}\to 0 \quad {\rm as} \quad k\to \infty
	$$
Let us concentrate on the decay of $\Phi$ operator.
We examine how the principle operator behaves in order to check the validity of the pseudo-resolvent condition mentioned above.
	\begin{align} 
\nonumber	\Phi(E) &=  \left.  (H_0 - E + \mu_P) \bigg[  1 +\underbrace{\lambda^2 \lip \ipt  \frac{1}{2p} \frac{1}{(\lop- \mu_P)(H_0-E+\lop)}}_{\tilde K} \right.   
 \\
	& \qquad - \left. \underbrace{\lambda^2 \lip\ipt\liq\iqt \frac{1}{2 \sqrt{pq} } \frac{1}{H_0-E+\mu_P}  \ladq \frac{1}{H_0 - E  + \lop + \loq} \lap}_{\tilde U} \bigg] \right .
	\end{align}
	The idea is this: for large negative values of $E$, we can make $||\tilde K(E)||<1/4$ as well as $||\tilde U(E)||<1/4$
thus we have the estimate,
\begin{eqnarray}
|\lambda_k| ||\Phi^{-1} (-|\lambda_k|)||&\leq& |\lambda_k| ||(H_0+|\lambda_k|+\mu_P)^{-1} ||\, ||[1+\tilde K(-|\lambda_k|)-\tilde U(-|\lambda_k|)]^{-1}||]\nonumber\\
&\leq& |\lambda_k| ||(H_0+|\lambda_k|+\mu_P)^{-1}||{1\over 1-||\tilde K(-|\lambda_k|)||-||\tilde U(-|\lambda_k|)||}\leq2|\lambda_k| ||(H_0+|\lambda_k|+\mu_P)^{-1}||\nonumber
.\end{eqnarray}
This implies that as $k\to\infty$, $|\lambda_k| ||\Phi^{-1} (-|\lambda_k|)||$ remains bounded.

To establish the above claims, let us look at the behavior of $\tilde K$ for 	real values of $E$. Later we  actually estimate these norms for complex values, there is a certain degree of repetition here but it is nice to see the differences,  a more proper thing to do is to use norm inequalities (since we do not have a formal proof that the operators are self-adjoint). 
	\begin{align}
	&\|\tilde K(E)\| = \lambda^2\Big\| \lip \ipt \frac{2p}{(-2p \mu_P + p^2 + m^2 + \pt^2)[(H_0-E)2p +p^2 +m^2 + \pt^2]}\Big\| \nonumber\\
	&= \frac{\pi}{2} \lambda^2\Big\| \lip \frac{2p}{(-2p \mu_P + p^2 + m^2 )[(H_0-E)2p +p^2 +m^2 ]^{1/2} + (-2p\mu_P +p^2+m^2)^{1/2} [(H_0-E)2p +p^2 +m^2 ]}\Big\|\nonumber\\
	&=  \frac{\pi}{2} \frac{\lambda^2}{2\pi} \int_0^\infty \frac{2p dp}{(p^2+m^2-2p \mu_P  )^{1/2}[(nm-E)2p +p^2 +m^2 ]^{1/2} (  (-2p\mu_P +p^2+m^2)^{1/2} + [(nm-E)2p +p^2 +m^2 ]^{1/2})} \nonumber\\
	& \leq \frac{\pi}{2} \lambda^2 \lip \frac{2p}{2(-2p \mu_P + p^2 + m^2 )[(nm-E)2p +p^2 +m^2 ]^{1/2} }
	\end{align}
Where in the third  line we replaced $H_0$ with its lower bound $nm$, for the last inequality we use the fact that $-\mu_P \leq nm-E$  we replace the bigger term $(nm-E)2p+m^2+p^2$ with the smaller term $-2p\mu_p+p^2+m^2$ to get an upper bound.

Now using Feynman parametrization  we get:
\begin{align}
 ||\tilde K(E)||&\leq C_1 \frac{\pi}{2} \lambda^2 \lip \int_{0}^{1}  \frac{p (1-u)^{-1/2}du}{[(nm-E-(nm-E+\mu_P)u)2p +p^2 +m^2]^{3/2} }\\
 &\leq C_1 \frac{\pi}{2} \lambda^2  \int_{0}^{1}   (1-u)^{-1/2}du  \bigg[ \frac{1}{(n+1)m-E-u(nm-E+\mu_P)} \bigg] \\
 &\leq C_2\lambda^2 \int_0^1  v^{-1/2}dv \bigg( \frac{1}{ (m-\mu_P) + (nm+\mu_P-E)v} \bigg) \\
 &=C_2\lambda^2 \int_0^1 d\eta \frac{1} {(m-\mu)+(nm+\mu_P-E)\eta^2}  
   \xrightarrow[E \to -\infty]{} 0 \nonumber
\end{align}
 This establishes that $||\tilde K(E)||$ goes to zero as $E\to -\infty$. Moreover, it shows that the operator remains bounded as long as $E<nm+\mu_P$, since the expression in (40) is well defined within this region, which is important for our later purposes. In fact, we can assume that the variable $E$ is complex, and the region of interest becomes $\operatorname{Re}(E)<nm+\mu_P$. Indeed, in the above  inequalities we may replace $E$ with its real part as long as we use it as a norm inequality, that gives an upper bound. If we further keep the imaginary part, we see that there is never a problem but we do not need this fact for our purpose (indeed we can extend the region of validity of the formulae by keeping a nonzero imaginary part).
 
So far we have shown that both $||\tilde U||$ (from eq. (\ref{eq: U bound}) it is obvious) and $||\tilde K||$ go to zero as $E$ goes to infinity. This indicates that $|\lambda_k | ~ || \Phi ^{-1} ( - |\lambda_k | ) ||$
 remains finite (this estimate is required for the  $\delta$ term as well).

 \textbf{$\beta$ term}: We have the following upper bound,
\begin{eqnarray}
\Big\|  |\lambda_k| \Phi^{-1} (-|\lambda_k|) \phi^{(+)} \frac{1}{H_0 + |\lambda_k|} \ket{f^{n+1}}  \Big\| \leq \underbrace{|\lambda_k| ~ || \Phi^{-1} ||}_\text{finite} ~ || \phi^{(+)}   \frac{1}{H_0 + |\lambda_k|} \ket{f^{n+1}} || ~~~~~~ 
\end{eqnarray}
Therefore we need to estimate the second norm,
\begin{align}
& \Big\|  \lip \ipt \frac{\lap}{\sqrt{2p}} \frac{1}{H_0+|\lambda_k|}  \ket{f^{n+1}} \Big\|
 =\Big\|  \lip \ipt \frac{1}{\sqrt{2p}} \frac{1}{H_0+|\lambda_k|+\lop} \lap \ket{f^{n+1}} \Big\|  \\
& \leq   \Big\|  \lip \ipt \underbrace{\frac{1}{\sqrt{2p}} \frac{1}{nm+|\lambda_k|+\lop}}_{g} a(p, p_\perp)\ket{f^{n+1}} \Big\| \\
\end{align} 
We now recall the inequality  
\bb
||\phi^{(+)}(g)\ket{f^n}||\leq n ||g||||f^n||
\ee
where all the norms are in the Hilbert space, which can be proven similar to the integral operator version. We have
\begin{equation}
 ||g||  = \frac{1}{2\pi} \bigg[ \lip \ipt \frac{1}{2p} \frac{1}{(nm+|\lambda_k|+\frac{p^2+\pt^2+m^2}{2p})^2} \bigg]^{1/2}
\end{equation}
Evaluating the integral we get the final result for $\beta$:
\bb
\Big[ \frac{nm + |\lambda_k|}{(nm)^2 +|\lambda_k|^2-4m^2} - \frac{4m}{(nm+|\lambda_k|)^2 -4m^2} \Big]  \xrightarrow[|\lambda_k| \to \infty]{} 0
\ee
Here $\alpha$ term requires a bit more work but 
 $\alpha, \gamma, \delta$  terms all go to zero as $|\lambda_k|$ goes to infinity in a similar way.
 
\textbf{$\alpha$ term}:
We start with the inequality:
\begin{eqnarray}
\!\!\!\!\!\!\!\!\Big\|  \Big[ |\lambda_k| \alpha( -|\lambda_k| ) - 1 \Big] \ket{f^{n+1}} \Big\| & \leq & \Big\| ~ \Big[ \frac{|\lambda_k|}{H_0 + |\lambda_k|} - 1 \Big] \ket{f^{n+1}} ~ \Big\|  \\ 
\nonumber
&+& ~ |\lambda_k| ~ || \Phi^{-1} || ~ || \frac{1}{H_0 + |\lambda_k|} \phi^{(-)} || ~ ||\phi^{(+)} \frac{1}{H_0 + |\lambda_k|} ~ \ket{f^{n+1}} ||
\end{eqnarray}
Notice that the first term involving only the resolvent $(H_0+|\lambda_k|)^{-1}$  actually goes to zero. Let us therefore concentrate on the next piece,
\begin{align}
\nonumber
&    \underbrace{ |\lambda_k| \| \Phi^{-1} \|}_\text{finite}
|| \frac{1}{H_0 + |\lambda_k|} \phi^{(-)} || ~ \underbrace{||\phi^{(+)} \frac{1}{H_0 + |\lambda_k|} ~ \ket{f^{n+1}}||}_{\rightarrow 0}
\end{align}
It is enough to look at,
\bb
\lim_{|\lambda_k| \rightarrow \infty} \Big\| \frac{1}{H_0 + |\lambda_k|} \lip \ipt \ladp \frac{1}{\sqrt{2p}} \ket {f^n}\Big\| \leq 
\lim_{|\lambda_k| \rightarrow \infty} \Big\| \lip \ipt \frac{\ladp}{\sqrt{2p}} \frac{1}{nm + |\lambda_k|+\omega(p,p_\perp)}\ket {f^{n+1}}\Big\| \nonumber
\ee
This factor behaves much like the preceding  factor, by means of a similar inequality for the creation part. Thus, we get $\lim_{|\lambda_k| \rightarrow \infty} \Big\|  \Big[ |\lambda_k| \alpha( -|\lambda_k| ) - 1 \Big] \ket{f^{n+1}} \Big\|   = 0$

\textbf{$\gamma$ term}:
The $\gamma$ term is identical to the previous expression it is the formal adjoint, written explicitly
\begin{eqnarray}
\nonumber
\Big\| |\lambda_k| \frac{1}{H_0 + |\lambda_k|} \lip \ipt \frac{\ladp}{\sqrt{2p}} \Phi^{-1}(-|\lambda_k|) \ket{f^n} \Big\| \leq |\lambda_k| ~ || \Phi^{-1} || ~ || \frac{1}{H_0 + |\lambda_k|} \lip \ipt \frac{\ladp}{\sqrt{2p}}  \ket{ f^n} ||
\end{eqnarray}
and by the previous arguments,
\bb
\lim_{|\lambda_k| \rightarrow \infty} || ~|\lambda_k| \gamma(-|\lambda_k|) ~ \ket{f^n} ~ || = 0
\ee
\textbf{$\delta$ term}:  
 Let us briefly mention  the $\delta$ term. The difference can be written as
 \begin{eqnarray}
 &\ & \Big\|\Big[|\lambda_k|\Phi^{-1}(-|\lambda_k|)-1\Big]\ket{f^n} \Big\|\leq \Big\| \Big[|\lambda_k|(H_0+|\lambda_k|+\mu)^{-1} -1\Big] \ket{f^n}\Big\|\nonumber\\
 &\ & \quad \quad \quad \quad \quad \quad ~~~~~~~~~~~~~~~~~~~~~~~~ +\underbrace{\Big\||\lambda_k|(H_0+|\lambda_k|+\mu)^{-1}\Big\|}_{\rm finite}\Big[ ||\tilde K||+||\tilde U||\Big] \frac{1}{1-||\tilde K||-||\tilde U||}
 \end{eqnarray}
 The finiteness of term as indicated above is due to the principle of uniform boundedness. Thus the whole expression goes to zero again using the above estimates.

\section{Common domain of of the family $\Phi(E)$ in light-front coordinates}\label{common_domain_lf}
In order to show the common domain of the family, we need to show the boundedness of the resolvent. We start by defining the kinetic and the potential terms as such, 
\begin{align}
	&\Phi(E)  =  \left.  \bigg[  1 +\underbrace{\lambda^2 \lip \ipt  \frac{1}{2p} \frac{1}{(\lop- \mu_P)(H_0-E+\lop)}}_{\tilde K} \right.   \nonumber\\
	& \qquad - \left. \lambda^2 \underbrace{\lip\ipt\liq\iqt   
	\frac{\ladq}{\sqrt{2q}} \frac{1}{H_0-E+\loq+\lop}   \frac{1}{H_0 - E +\mu_p + \lop } \frac{\lap}{\sqrt{2p}}}_{\tilde U}  \bigg]\right.\nonumber\\
	& \left. \quad \quad \quad \quad \quad \quad \quad \quad \quad \quad \quad \quad \quad \quad \quad \quad \quad \quad \quad \quad~~~~~~~~~~~~~~~~~~~~~~~~~~~~~~~~~~ \times (H_0 - E + \mu_P)\right. 
	\end{align}
	Let us now work out a norm estimate on the $\tilde U$ term since we already commented on the $\tilde K$ term above.
	We use a slightly different approach since our aim is to show that this term is  bounded as long as $\operatorname{Re}(E)<nm+\mu$
	We remind again the inequality
	$$
	|| \int dpdq a^\dag (p) F(p,q) a(q)||< n \Big [ \int dpdq |F(p,q)|^2 \Big ]^{1/2}
	$$
	where the integral refers to multi-parameters and norm is taken in a Fock space of $n$ particles.
	This implies immediately that if we drop  $\operatorname{Im}(E)$ we get an upper bound, so we replace $E$ with $\operatorname{Re}(E)$. Moreover inside the norm we replace the positive operator $H_0$ by  its lower bound $(n-1)m$,
	\begin{align}
	& \|\tilde U\|= \lambda^2 \Big\| \lip \ipt \liq \iqt \frac{a^\dagger(q, q_\perp)}{\sqrt{2p}}\nonumber\\
	&\hskip 3 cm \times \frac{1}{[H_0-E +\omega(p,p_\perp)+\omega(q,q_\perp)][H_0+\mu_p-E+\omega(p,p_\perp)] }
	\frac{a(p,p_\perp)}{\sqrt{2q}}\Big\|\nonumber \\
	& \hskip 1 cm \leq \lambda^2 n \Bigg[ \int_0^\infty \frac{dpdq}{4\pi^2}  \int \frac{dp_\perp dq_\perp}{ 4\pi^2} \frac{1}{{2p}}
	\frac{1}{[(n-1)m-\operatorname{Re}(E) +\omega(p,p_\perp)+\omega(q,q_\perp)]^2[(n-1)m+\mu_p-\operatorname{Re}(E)+\omega(p,p_\perp)]^2 }
	\frac{1}{2q}\Bigg]^{1/2}\nonumber 
	.\end{align}
	Let us now note that $(n-1)m-\operatorname{Re}(E)=nm+\mu_p-\operatorname{Re}(E)+2{1\over 2}(m+\mu_p)$ and let us suppose we always keep $\operatorname{Re}(E)<nm+\mu_p$, which is the region of interest for possible bound states. 
	We now use a generalized arithmetic-geometric mean inequality (for positive numbers),
	$$
	  \omega(p,p_\perp)-{1\over 2} (m+\mu_p) +\omega(q,q_\perp)-{1\over 2} (m+\mu_p)> [\omega(p,p_\perp)-{1\over 2} (m+\mu_p)]^{1/4} [\omega(q,q_\perp)-{1\over 2} (m+\mu_p)]^{3/4}
	  .$$
	  This splits the integration. We drop the difference $\Delta=nm+\mu_p-\operatorname{Re}(E)$ for simplicity in ($q,q_\perp$)-integrals. Furthermore 
	  we write $\omega(p,p_\perp)+(n-1)m+\mu_p-\operatorname{Re}(E)$ term as $\omega(p,p_\perp)-(m-\Delta)$, and combine the two ($p,p_\perp$)-terms
with the largest of ${m+\mu_p\over 2}$	and $m-\Delta$, calling the largest one $m_*$, which is strictly less than $m$. After cancelling the $p$ and $q$ products, we have the product of two integrals below,
\begin{eqnarray}
\| \tilde U \|&\leq& C_1 \lambda^2 \Bigg[ \int_0^\infty \int \frac{p^{3/2}dpdp_\perp}{ [p^2+p_\perp^2+m^2 -2m_*p]^{5/2}}\int_0^\infty\int 
\frac{q^{1/2} dqdq_\perp}{[q^2+q_\perp^2 +m^2 -(m+\mu_p)q]^{3/2} }\Bigg]^{1/2}\nonumber\\
 &\leq& C_2 \Bigg[\int_0^\infty\rho^{5/2} d\rho\int_{-\pi/2}^{\pi/2}d\theta \frac{\cos^{3/2}(\theta)}{[\rho^2+m^2-2m_*\rho\cos(\theta)]^{5/2}}\int_0^\infty\tau^{3/2} d\tau\int_{-\pi/2}^{\pi/2}d\beta \frac{\cos^{1/2}(\beta)} {[\tau^2 +m^2-(m+\mu_p)\tau \cos(\beta)]^{3/2}}\Bigg]^{1/2}\nonumber,
\end{eqnarray}
both of which are finite integrals. Consequently, we have the desired result for complex $E$ for which $\operatorname{Re}(E)<nm+\mu_p$.
The holomorphicity requirement will make use of this bound as well. 
Although the above argument can be generalized to discuss the decay of $\tilde U$ for large negative values of 
$\operatorname{Re}(E)$, it is instructive to get another bound by means of Feynman parametrization and exponentiation. Therefore we give an alternative estimate for the norm,
	\begin{align}
	    &\| \tilde U\| = \Bigg\| \lambda^2 \lip\ipt\liq\iqt \frac{1}{2 \sqrt{pq}}  \ladq \int_{0}^{1}   \frac{du}{
	    (\loq u +(1-u)\mu_P+H_0-E+\lop)^2}    \lap \Bigg\| \\
	    &= \lambda^2 \Bigg\| \lip\ipt\liq\iqt \frac{1}{2 \sqrt{pq}}  \int_{0}^{1} du 
	    \int_{0}^{\infty} sds e^{-s(1-u)\mu_P}  \ladq e^{-s(\loq u  +H_0-E+\lop)}  \lap  \Bigg\|
	\end{align}
For ease of calculation let us define:	
\begin{eqnarray}
\nonumber
\phi^{(+)} (h(s)) &=&  \lip\ipt \frac{1}{\sqrt{p}} \lap e^{-s\lop}  \\ 
\phi^{(-)} (g(su)) &=&  \liq\iqt \frac{1}{\sqrt{q}} \ladq e^{-su\loq}
\end{eqnarray}
Then the norm inside $\tilde U$ can be estimated as follows, after pulling out $s$ and $u$ integrals, $e^{sE}$ is  replaced with $e^{s\operatorname{Re}(E)}$,
moreover we  replace  $H_0$ by its lower bound $nm$.  We also use the inequality,
\begin{align}
     \| \phi^{(-)}(g(su)) e^{-sH_0}  \phi^{(+)} (h(s)) \| \leq e^{-s(n-1)m} n \| g \| \| h\|  
\end{align}
Then  the norm becomes,
\begin{align}
   & \| \phi^{(-)}  \phi^{(+)} \| \leq  n \Bigg[\lip \ipt \frac{1}{{p}} e^{-su \frac{1}{p} (p^2+m^2+\pt^2)}\Bigg]^{1/2}
    \Bigg[\liq\iqt \frac{1}{q} e^{-s \frac{1}{q}(q^2+m^2+\qt^2)} \Bigg]^{1/2}\\
   & \quad \quad \quad \quad \quad \quad  \leq C_1 n \frac{e^{-s(1+u)m/2} } {u^{1/2} s}
\end{align}
Here we use the integral 
\begin{equation}
    \lip {1 \over \sqrt{p}} e^{-s(p+\frac{m^2}{p}) } = {1\over 2\sqrt{\pi s}  }e^{-ms}  
\end{equation}
Putting back this into the inequality for $\tilde U$:
	\begin{align}
	    &\|\tilde U\| \leq C_1n\lambda^2    \int_{0}^{1} du  \int_{0}^{\infty} sds e^{-s(1-u)\mu_P-s(n-1)m-sRe(E)} \frac{e^{-s(1+u)m/2} } { s} \\
	    & \leq C_2\lambda^2 n \int_0^\infty ds e^{s[(n-1)m+\mu_p -\operatorname{Re} (E)]} \int_0^1 {du\over u^{1/2}} e^{-su(m-\mu_p)-s(1-u)m/2}  \\
	    & \leq C_3 {\lambda^2 n\over (n-1)m+\mu_p-\operatorname{Re}(E)} 
	\end{align}
 Thus, both of the operators are bounded.

\section{Verifying the decay conditions on the manifold}\label{resolvent-manifold}

 Noting that the free resolvent $ R_0 = \frac{1}{H_0 - E} $ satisfies (\ref{eq: reseq}), it is straightforward to show that (\ref{eq: isres}) reduces to:
\begin{eqnarray}
R_0(E_1) b^\dagger \Phi^{-1}(E_2) ~ \big[ \Phi(E_1) - \Phi{(E_2)} + b \Big( R_0(E_1) - R_0(E_2) \Big) 
b^\dagger + E_1 - E_2 \big] ~ \Phi^{-1}(E_2) b R_0(E_2) \overset{?}{=} 0
\label{eq: reslast}
\end{eqnarray}
We can check the equality in (\ref{eq: reslast}) by direct substitution. Calculating the term in square brackets term by term:
\begin{eqnarray}
\nonumber
\mathcal{A} = \Phi(E_1)-\Phi(E_2) &=& (H_0 - E_1 +\mu_p) - (H_0 -E_2 + \mu_p ) \\
 &+& \lambda^2 \sum_\sigma \frac{|f_\sigma | ^2}{2 \omega_\sigma} \frac{E_2-E_1}{(H_0-E_1+\omega_\sigma)(H_0-E_2+\omega_\sigma)}
\label{eq: phidif} \\
\nonumber
&+& \lambda^2  \sum_{\sigma,\tau} f_\sigma (\bar{x}) \frac{a_\sigma ^\dagger}{\sqrt{2\omega_\sigma}} \frac{E_2-E_1}{(H_0-E_1+\omega_\sigma+\omega_\tau)} 
\frac{1}{(H_0-E_2+\omega_\sigma+\omega_\tau)} \frac{a_\tau}{\sqrt{2\omega_\tau}}
\end{eqnarray}

\begin{eqnarray}
\nonumber
\mathcal{B} = b \Big( R_0(E_1) - R_0(E_2) \Big) 
b^\dagger &=& 
(\lambda \sum_\sigma \frac{f_\sigma (\bar{x})}{\sqrt{2 \omega_\sigma}} a_\sigma ) \big[ \frac{1}{H_0 - E_1} - \frac{1}{H_0 - E_2} \big]
(\lambda \sum_\tau \frac{f_\tau (\bar{x})}{\sqrt{2 \omega_\tau}} a_\tau ^\dagger ) \\
\nonumber
&=& \frac{\lambda^2}{2} \sum_{\sigma,\tau} \frac{f_\sigma f_\tau}{\sqrt{\omega_\sigma \omega_\tau}} (E_1-E_2) \frac{1}{H_0 - E_1 + \omega_\sigma} a_\sigma a_\tau ^\dagger \frac{1}{H_0 - E_2 + \omega_\tau} \\
\nonumber
 &=& \frac{\lambda^2}{2} \sum_{\sigma,\tau} \frac{f_\tau a^\dagger _\tau}{\sqrt{\omega_\tau}} \frac{E_1-E_2}{(H_0 - E_1 + \omega_\sigma + \omega_\tau)}\frac{1}{(H_0 -E_2 +\omega_\sigma + \omega_\tau)}  \frac{f_\tau a_\tau}{\sqrt{\omega_\tau}} \\
\nonumber
&+& \frac{\lambda^2}{2} \sum_\sigma \frac{|f_\sigma|^2}{\omega_\sigma} \frac{E_1-E_2}{(H_0-E_1+\omega_\sigma)(H_0-E_2+\omega_\sigma)}\\
\nonumber
\mathcal{C} = E_1 - E_2~~~~~~~~~~~~~~~~~~~~ &\ &   ~~~~~~~~~~~~~~~~~~~~~~~~~~~~~~~~~~~~~~~~~~~~~~~~~ \\
\end{eqnarray}
As we can see now, we have, 
\bb
\mathcal{A} + \mathcal{B} + \mathcal{C} = 0
\ee
Thus, $R(E)$ is indeed a pseudo-resolvent family that depends  on  a { \it complex} parameter $E$.

\textbf{The Decay Condition:}
\label{subsection: decay}
To show that the resolvent $R$ satisfies Theorem \ref{thm: decay}, as before, we choose a series $\lambda_k$ on the negative real axis such that for every $k$, $\lambda_k <0< E_{gr}$. Since $\lambda_k = -|\lambda_k|$, we can write down the condition (\ref{eq: decaycond}) as:
\bb
\lim_{k \rightarrow \infty} \Big| \Big| [|\lambda_k| R(-|\lambda_k|) -1] x \Big| \Big| _{\mathcal{H}}  = 0
\label{eq: deccon}
\ee
Substituting the resolvent expression in the previous equation, and 
applying the triangular inequality twice, we again conclude that,
\begin{eqnarray}
&\ &\Big\| \Big[ |\lambda_k| R(-|\lambda_k|) -1 \Big] \begin{pmatrix}
\ket{f^{n+1}} \\
\ket{f^n}
\end{pmatrix}
\Big\| \leq 
\nonumber   \\
&\ & || ~ \big( |\lambda_k| \alpha(-|\lambda_k|) -1 \big)\ket{f^{n+1}} ~ || + ||~ |\lambda_k| ~ \gamma(-|\lambda_k|) \ket{f^n}||
\label{eq: line1}  + || ~|\lambda_k| \beta(-|\lambda_k|) ~ \ket{f^{n+1}} || + ||~ \Big( |\lambda_k| \delta(-|\lambda_k|)-1 \Big) \ket{f^n} || \nonumber\\
\end{eqnarray}
So, if we can show that as $k \rightarrow \infty$, each term in (\ref{eq: line1})  goes to zero separately, (\ref{eq: deccon}) can be immediately deduced.

\textbf{Behaviour of the $\Phi$ Operator:}
Since we need to reconsider  the behaviour of the operator for complex values of $E$, the estimates below use  $E$ as a complex variable. Let us write the operator $\Phi$ below, as before, removing the operator $H_0-E+\mu_p$ to the left.
\begin{eqnarray}
\nonumber 
 \Phi(E) &=& (H_0 - E + \mu_p ) ~ \big\{ 1 + ~ \underbrace{\frac{\lambda^2}{2} \sum_\sigma f_\sigma ^2 (\bar{x}) \frac{1}{\omega_\sigma (\omega_\sigma - \mu_p) ( H_0 -E + \omega_\sigma)}}  \\
\nonumber
&\ & ~~~~~~~~~~~~~~~~~~~~~~~~~~~~~~~~~~~~~~~~~~~~~~~~~ K(E) \\
\nonumber & \ & \\
\nonumber
 &-& \underbrace{ \frac{\lambda^2}{2} \sum_{\sigma , \tau} f_\sigma(\bar{x}) f_\tau (\bar{x}) \frac{1}{H_0 - E + \mu_p} \frac{a_\sigma ^\dagger}{\sqrt{\omega_\sigma}} \frac{1}{H_0 - E + \omega_\sigma + \omega_\tau} \frac{a_\tau}{\sqrt{\omega_\tau}} } \Big\} ~~~~~~~~\\
\nonumber
&\ & 
~~~~~~~~~~~~~~~~~~~~~~~~~~~~~~~~~~~~ U(E)
\end{eqnarray}
To understand the behaviour of $\Phi$ and make use of it in the upcoming calculations, we estimate $K$ and $U$ here. Applying Feynman parametrization to $K$:
\begin{eqnarray}
\nonumber
K &=& \mathcal{C} \sum\limits_\sigma f_\sigma ^2(\bar x) ~   \int _0 ^1 du_1  du_2  du_3 \frac{\delta ( 1-\sum\limits_{i=1}^3 u_i)}{[ u_1 \omega_\sigma + u_2 (\omega_\sigma -\mu_p) + u_3 (\omega_\sigma + H_0 - E)] ^3} \\
&=& \mathcal{C} \sum\limits_\sigma  f_\sigma ^2(\bar x) ~ \int_0 ^1 du_1 du_2 du_3 ~ \delta( 1-\sum\limits_{i=1}^3 u_i) \int s^2 ds e^{-sA}
\end{eqnarray} 
where all constants are absorbed into $\mathcal{C}$ and we define as $ A =  u_1 \omega_\sigma + u_2 (\omega_\sigma -\mu_p) + u_3 (\omega_\sigma + H_0 - E) $. Note that because of  the Dirac delta function, $ u_1+u_2+u_3 = 1$  and we have:
\bb
K = \mathcal{C}  ~ \int_0 ^1 du_1 du_2 du_3 ~ \delta( 1-\sum\limits_{i=1}^3 u_i) \int s^2 ds [\sum\limits_\sigma  f_\sigma ^2(\bar x)e^{-s \omega_\sigma}] e^ {s\mu_p u_2} e^{-s u_3(H_0 -E)}
\ee
Now applying the subordination identity to $e^{-s\omega_\sigma}$ and substituting the heat kernel:
\begin{equation}
K = \mathcal{C}  \int_0 ^1 du_1 du_2 du_3  \delta( 1-\sum\limits_{i=1}^3 u_i) 
\int s^3 ds \int d\xi  \frac{e^{-\frac{s^2}{4 \xi} -m^2 \xi}}{\xi^{3/2}} K_\xi (\bar{x},\bar{x}) e^ {s\mu_p u_2} e^{-s u_3(H_0 -E)}
\label{eq: Kforint}
\end{equation}
If we substitute the estimate for the heat kernel (\ref{eq: hkestimate}) and take the norm:
\bb
|| K || \leq \mathcal{C} \int s^3 ds du_2du_3  e^{su_2 \mu_p} e^{-s(nm - \operatorname{Re}(E))u_3} 
\int d\xi \frac{m  e^{-\frac{s^2m^2}{4 \xi}} e^{-\xi}}{\xi^{3/2}}  \Big[ \frac{C}{\xi / m^2} + \frac{1}{V(\mathcal{M})} \Big]
\label{eq: K}
\ee
where we estimate $H_0$ as $(nm)$ and note that $E \rightarrow \operatorname{Re}(E)$ when we use  the norm. The two parts of the integral is examined separately, we call them $||K||_{(1)}$ and $||K||_{(2)}$. Looking at the first part ($\frac{C}{\xi/m^2}$ term), this is the integral representation of the modified Bessel function of the second kind:
\bb
K_\nu (z) = \frac{1}{2} (\frac{1}{2} z )^\nu \int _0 ^\infty \frac{1}{\xi^{\nu +1}} e^{-\xi - \frac{z^2}{4\xi}} d\xi
\ee
where in our case $z=ms$ and $\nu=3/2$. Then the integral can be computed to get:
\bb
K_{3/2} (ms) = \sqrt{\frac{\pi}{2ms}} e^{-ms} \Big[ 1 + \frac{1}{ms} \Big]
\ee
we arrive at:
\bb
||K||_{(1)} \leq \mathcal{C} \int ms ds du_2du_3  e^{su_2 \mu_p} e^{-s(nm - \operatorname{Re}(E))u_3}  e^{-ms} \Big[ 1 + \frac{1}{ms} \Big]
\ee
where we absorb every constant into $\mathcal{C}$. If we find  an upper bound to the most singular part of the integral  coming from the estimate of the Bessel function (i.e. the second term in square brackets), one could see that a bound for the other part is easier to obtain. As we take the limit $E=\lambda_k \rightarrow - \infty$, the bound we are  to find is enough for our  calculations below as well. Denote the norm coming from the most singular part as $||K||_{(1)-sing}$. Introducing $u_1 + u_2 + u_3 = 1 $:
\begin{eqnarray}
\nonumber 
||K||_{(1)-sing} &\leq & \mathcal{C} \int ds du_1du_2du_3 \delta( 1-\sum\limits_{i=1}^3 u_i) e^{su_2 \mu_p} e^{-s(nm - \operatorname{Re}(E))u_3}  e^{-ms(u_1 + u_2 + u_3)} \\
\nonumber
&\leq & \mathcal{C} \int ds du_1du_2du_3 \delta( 1-\sum\limits_{i=1}^3 u_i) e^{-smu_1} e^{-s(m-\mu_p)u_2} e^{-s\big( (n+1)m- \operatorname{Re}(E) \big) u_3} \\
\nonumber
&\leq & \mathcal{C} \int  du_1du_2du_3 \delta( 1-\sum\limits_{i=1}^3 u_i) 
\frac{1}{mu_1 + (m-\mu_p)u_2 + [(n+1)m- \operatorname{Re}(E)]u_3}
\end{eqnarray}
If we ignore  the term $(m-\mu_p)u_2$ and take the $u_2$ integral:
\bb
||K||_{(1)-sing} \leq  \mathcal{C}  \int\limits_{0 \leq u_1 + u_3 \leq 1} \frac{du_1 du_3}{mu_1 + [(n+1)m-\operatorname{Re}(E)]u_3 }
\ee
Note that the region $0\leq u_1 + u_3 \leq 1$ is contained in $u_1 ^2 + u_3 ^2 \leq 1$, so we can integrate in the latter since the integrand is positive. We can go to polar coordinates where $u_1 = \rho \cos \theta$ and $u_3 = \rho \sin \theta$:
\bb
||K||_{(1)-sing} \leq  \mathcal{C}  \int\limits_{0} ^{1} \int\limits _{0} ^{\pi/2} \frac{\rho d\rho d\theta}{m\rho \cos \theta  + [(n+1)m - \operatorname{Re}(E)]\rho \sin \theta}
\ee
Since in the first quadrant both sine and cosine are positive, we can write the inequalities $\cos \theta \geq \cos^2 \theta$ and $\sin\theta \geq \sin^2 \theta$. We then replace  cosine and sine with the squares and turn the integral to a more familiar form:
\begin{eqnarray}
\nonumber
||K||_{(1)-sing} &\leq &  \mathcal{C}  \int\limits_{0} ^{1} \int\limits _{0} ^{\pi/2} \frac{d\rho d\theta}{m \cos^2 \theta + [(n+1)m - \operatorname{Re}(E)] \sin^2 \theta} 
\leq  \mathcal{C} ~\frac{\pi}{\sqrt{m}} \frac{1}{ \sqrt{(n+1)m - \operatorname{Re}(E)}}
\end{eqnarray}
Thus in the limit 
\bb
E = \lambda_k \rightarrow -\infty ~~ , ~~ 
||K||_{(1)} \rightarrow 0
\label{eq: k1lim}
\ee
For $||K||_{(2)}$, we proceed similarly:
\begin{equation}
|| K ||_{(2)} = \mathcal{C} \int s^3 ds du_1du_2du_3 \delta( 1-\sum\limits_{i=1}^3 u_i) e^{su_2 \mu_p} e^{-s(nm - \operatorname{Re}(E))u_3} 
\int d\xi  \frac{m ~ e^{-\frac{s^2m^2}{4 \xi}} e^{-\xi}}{\xi^{3/2}} ~ \Big[  \frac{1}{V(\mathcal{M})} \Big] 
\end{equation}
\begin{eqnarray}
\nonumber 
||K||_{(2)} &\leq &  \frac{\mathcal{C}}{V(\mathcal{M})} \int ds  du_2 du_3  s^{5/2}e^{su_2 \mu_p} e^{-s(nm - \operatorname{Re}(E))u_3}  \frac{e^{-ms}}{\sqrt{ms}} \\
\nonumber
& \leq &  \frac{\mathcal{C}}{V(\mathcal{M})} \int ds du_1 du_2 du_3\delta( 1-\sum\limits_{i=1}^3 u_i) s^2 e^{-s [ mu_1 + (m-\mu_p)u_2 + ((n+1)m- \operatorname{Re}(E))u_3]} \\
\nonumber
&\leq & \frac{\mathcal{C}}{V(\mathcal{M})} \int\limits _0 ^1 du_1 du_2 du_3\delta( 1-\sum\limits_{i=1}^3 u_i)
\frac{1}{\big[ mu_1 + (m-\mu_p)u_2 + \big( (n+1) m - \operatorname{Re}(E) \big) u_3 \big] ^3} \\
\nonumber
&\leq & \frac{\mathcal{C}}{V(\mathcal{M})} \int\limits _0 ^1 du_3 \frac{1}{[ (m-\mu_p) + (nm + \mu_p - \operatorname{Re}(E))u_3 ]^{3}} \\
&\leq& 
\frac{\mathcal{C}}{V(\mathcal{M})} \Big[ \frac{1}{2 (nm + \mu_p - \operatorname{Re}(E)) (m-\mu_p)^2} \Big]
\label{eq: K2}
\end{eqnarray}
It is straightforward to see that:
\bb
E = \lambda_k \rightarrow -\infty ~~ , ~~ 
||K||_{(2)} \rightarrow 0
\label{eq: k2lim}
\ee
it follows that:
\bb
E = \lambda_k \rightarrow -\infty ~~ , ~~ 
||K|| \rightarrow 0
\label{eq: klim}
\ee
We also has the bound for $||U||$, see eq \ref{eq: ubound}. Substituting back $ \chi = nm - \operatorname{Re}(E)$, it is straightforward to take the limit and see that:
\bb
E = \lambda_k \rightarrow -\infty ~~ , ~~ 
||U|| \rightarrow 0
\label{eq: ulim}
\ee
We add one more result to this section for future simplicity.  Note that:
\bb
|\lambda_k | ~ || \Phi ^{-1} ( - |\lambda_k | ) || = |\lambda_k |~ || (1+ K(-|\lambda_k |) - U(- |\lambda_k | ))^{-1} || ~ ||~(H_0 + \mu_p + |\lambda_k | )^{-1} ||
\ee
When $ ||A|| < 1 $ , we can write down the Neumann series:
\bb
( 1- A)^{-1} = \sum\limits _{l=0} ^{\infty} A^l
\ee
Thus using (\ref{eq: klim}) and (\ref{eq: ulim}), we can deduce that for proper choice of $\lambda_k$ we can set:
\bb
||U||<\frac{1}{4} \quad {\rm and} \quad ||K||<\frac{1}{4}
\ee
Hence the Neumann series expansion leads to an upper bound: 
\begin{equation}
|| (1+ K(-|\lambda_k |) - U(- |\lambda_k | ))^{-1} || <2
\end{equation}
Consequently  we  deduce that,
\bb
|\lambda_k | ~ || \Phi ^{-1} ( - |\lambda_k | ) || \leq C |\lambda_k | ~ | (nm + |\lambda_k | ) ^{-1} |
\ee
It is now obvious that in the limit $ |\lambda_k | \rightarrow \infty$ the right hand side goes to a constant, which means  $|\lambda_k | ~ || \Phi ^{-1} ( - |\lambda_k | ) || $ is finite.

\textbf{ $\beta$ term:}
We can write the inequality, as such:
\begin{eqnarray}
 \Big\|  |\lambda_k| \Phi^{-1} (-|\lambda_k|) \phi^{(+)} \frac{1}{H_0 + |\lambda_k|} \ket{f^{n+1}}  \Big\| \leq |\lambda_k| ~ || \Phi^{-1} || ~ || \phi^{(+)}   \frac{1}{H_0 + |\lambda_k|} \ket{f^{n+1}} || ~~~~~~ \label{eq: betaineq}
\end{eqnarray}
without any problems since each term on the right hand side is bounded. We know that   $|\lambda_k| \Phi^{-1}(-|\lambda_k|)$ has finite  operator norm. We need to work on the following term:
\begin{eqnarray}
|| ~ \phi^{(+)} \frac{1}{H_0 + |\lambda_k|} \ket{f^{n+1}} ~ || = || ~ \sum_{\sigma}
\underbrace{\frac{f_\sigma}{\sqrt{2\omega_\sigma}} \frac{1}{H_0 + |\lambda_k| + \omega_\sigma} } a_\sigma \ket{f^{n+1}} ~ || \\ 
\nonumber
g (\sigma) ~~~~~~~~~~~~~~~~~~~~~~~~~
\end{eqnarray} 
Using the estimates as is done in the  previous part (substituting $H_0$ as $nm$ to get an upper):
\begin{eqnarray}
\nonumber
\Big\| \sum_\sigma \frac{f_\sigma}{H_0+ |\lambda_k| + \omega_\sigma} \frac{a_\sigma}{\sqrt{2\omega_\sigma}} \ket{f^{n+1}} \Big\| &\leq & (n+1) \Big[ \sum_\sigma \frac{|f_\sigma|^2}{(nm+ |\lambda_k| + \omega_\sigma)^2 \omega_\sigma} \Big]^{1/2} ||f^{n+1}|| \\
\nonumber
&\leq & (n+1) \Big[ \sum_\sigma \frac{|f_\sigma|^2}{(nm+|\lambda_k|)^2 + \omega_\sigma ^2} \frac{1}{\omega_\sigma} \Big] ^{1/2} ||f^{n+1}||
\end{eqnarray}
If we applies Feynman parametrization to the term in square brackets:
\begin{eqnarray}
\nonumber
& \ & \sum_\sigma  \frac{|f_\sigma|^2}{(nm+|\lambda_k|)^2 + \omega_\sigma ^2} \frac{1}{(\omega_\sigma ^2)^{1/2}} = \frac{1}{2} \int _0 ^1 \sum_\sigma \frac{d\xi (1-\xi)^{-1/2} |f_\sigma|^2} { {[\xi(nm+ |\lambda_k|)^2 + \omega_\sigma ^2 ] ^{3/2}} } \\
& \ &\quad \qquad \qquad \qquad   = \frac{1}{2} \int_0 ^1 \frac{d\xi}{\sqrt{1-\xi}} \int_0 ^\infty ds \sqrt{s} e^{-s\xi(nm+|\lambda_k|)^2} \underbrace{ \sum_\sigma |f_\sigma|^2 e^{-s \omega_\sigma ^2} } \label{eq: ab} \\
\nonumber
& \ & ~~~~~~~~~~~~~~~~~~~~~~~~~~~~~~~~~~~~~~~~~~~~~~~~~~~~~~~~~~~~~~~~~~~~~~~~~K_s (\bar{x},\bar{x}) e^{-sm^2} 
\end{eqnarray}
Substituting the estimate for the heat kernel (\ref{eq: hkestimate}) and computing the integrals we arrive at the following inequality:
\begin{eqnarray}
\nonumber \sum_\sigma  \frac{|f_\sigma(\bar x)|^2}{(nm+|\lambda_k|)^2 + \omega_\sigma ^2} \frac{1}{(\omega_\sigma ^2)^{1/2}}  &\leq & \frac{1}{2} \underbrace{ \int_0 ^1 \frac{d\xi}{\sqrt{\xi(1-\xi)}}} C \underbrace{\int_0 ^\infty ds  e^{-s^2(nm+|\lambda_k| )^2  }}  +  \frac{1}{2} \frac{1}{V(\mathcal{M})} \underbrace{\int_0 ^1 \frac{d \xi}{\sqrt{\xi (1-\xi)}}} \frac{1}{nm + |\lambda_k|}  \\ 
\nonumber & \ & ~~~~~~~~~~~\pi ~~~~~~~~~~~~~~~~~~~~~~~~\frac{1}{[nm+ |\lambda_k|]^{1/2}}~~~~~~~~~~~~~~~~~~~~~\pi \\
& \leq & \big[ \frac{C }{2 } + \frac{\tilde{C}}{V(\mathcal{M})} \Big] \frac{\pi}{[nm + |\lambda_k|]^{1/2}}
\label{eq: betalim}
\end{eqnarray}
where $\tilde{C}$ is a finite constant introduced for notational simplicity (which can be computed). Now we can substitute everything into (\ref{eq: betaineq}) to get:
\begin{eqnarray}
\Big\|  |\lambda_k| \Phi^{-1}  \phi^{(+)}  \frac{1}{H_0 + |\lambda_k|} \ket{f^{n+1}}  \Big\| & \leq & \underbrace{|\lambda_k| || \Phi ^{-1} ||} C \Big[ \frac{1}{nm+ |\lambda_k|} \Big] ^{1/4} ~~~~~~~ \label{eq: betasimplified} \\
\nonumber
& \ & ~~ {\rm finite} \label{eq: limit}
\end{eqnarray}
where we collect all constants (including the volume of the manifold) in $C$ for simplicity and make use of the fact that $\ket{f^{n+1}}$ is normalized. This implies directly that the left hand side of (\ref{eq: betasimplified})  goes to zero, exactly what we aimed to show:
\bb
\lim_{|\lambda_k| \rightarrow \infty} || ~|\lambda_k| \beta(-|\lambda_k|) ~ \ket{f^{n+1}} ~ || = 0
\label{eq: betafinal}
\ee

\textbf{ $\gamma$ term:}
We proceed as  in the previous part.
\begin{eqnarray}
\nonumber
\Big\| |\lambda_k| \frac{1}{H_0 + |\lambda_k|} \sum_\sigma \frac{f_\sigma(\bar x)}{\sqrt{2\omega_\sigma}} a^\dagger _\sigma \Phi^{-1}(-|\lambda_k|) \ket{f^n} \Big\| \leq |\lambda_k| ~ || \Phi^{-1}(-|\lambda_k|) || ~ ||\sum_\sigma \frac{1}{H_0 + |\lambda_k|} \frac{f_\sigma(\bar x)}{\sqrt{2\omega_\sigma}} a^\dagger _\sigma \ket{ f^n} ||
\end{eqnarray}
We consider the last factor, similar to previous estimates we see that:
\begin{eqnarray}
\nonumber
||\sum_\sigma \frac{1}{H_0 + |\lambda_k|}\frac{f_\sigma(\bar x) }{\sqrt{2\omega_\sigma}} a^\dagger _\sigma \ket{ f^n} ||  \leq
 \sqrt{n+1} \Big( &\sum \limits_\sigma & \frac{|f_\sigma|^2}{(nm+|\lambda_k|+\omega_\sigma)^2 \omega_\sigma} \Big) ^{1/2} || ~\ket{f^n}|| 
\label{eq: gammaa}
\end{eqnarray}
Note that on the right hand side, we have exactly what we have above up to some constant and we can use the result in (\ref{eq: limit}) directly to establish the result we seek after:
\bb
\lim_{|\lambda_k| \rightarrow \infty} || ~|\lambda_k| \gamma(-|\lambda_k|) ~ \ket{f^n} ~ || = 0
\label{eq: gammafinal}
\ee

\textbf{ $\alpha$ term}

We again start with an inequality:
\begin{eqnarray}
\Big\|  \Big[ |\lambda_k| \alpha( -|\lambda_k| ) - 1 \Big] \ket{f^{n+1}} \Big\| & \leq & \Big\| ~ \Big[ \frac{|\lambda_k|}{H_0 + |\lambda_k|} - 1 \Big] \ket{f^{n+1}} ~ \Big\|  \\ 
\nonumber
&+& ~ \underbrace{|\lambda_k| ~ || \Phi^{-1}(-|\lambda_k|) ||}_{\rm finite} ~ || \frac{1}{H_0 + |\lambda_k|} \phi^{(-)} || ~ ||\phi^{(+)} \frac{1}{H_0 + |\lambda_k|} ~ \ket{f^{n+1}} ||
\end{eqnarray}
Taking the limit, it is straightforward to see that the first term on the right hand side goes to zero. The second term should be worked out, in a similar way we  see that 
\begin{eqnarray}
\nonumber
\lim_{|\lambda_k| \rightarrow \infty} || \frac{1}{H_0 + |\lambda_k|} \phi^{(-)} \ket{f^n}||  = 0.~~~
\label{eq: supfimin}
\end{eqnarray}
Note that the last  term is exactly the same as we have above for $\gamma$ and so,
\bb
\lim_{|\lambda_k| \rightarrow \infty} ||\phi^{(+)} \frac{1}{H_0 + |\lambda_k|} ~ \ket{f^{n+1}} || \leq \lim_{|\lambda_k| \rightarrow \infty} C \Big(\frac{1}{(n+1)m+ |\lambda_k|}\Big) ^{1/4} = 0 
\label{eq: limalpha}
\ee
where $C$ is constant. Thus we conclude  that:
\begin{eqnarray}
\nonumber
\lim_{|\lambda_k| \rightarrow \infty}  \Big[ \Big\| ~ \Big[ \frac{|\lambda_k|}{H_0 + |\lambda_k|} - 1 \Big] \ket{f^{n+1}} ~ \Big\|  
 + |\lambda_k| ~ || \Phi^{-1}(-|\lambda_k|) || ~ || \frac{1}{H_0 + |\lambda_k|} \phi^{(-)} || ~ ||\phi^{(+)} \frac{1}{H_0 + |\lambda_k|} ~ \ket{f^{n+1}} || \Big] = 0 
\end{eqnarray}
Hence follows the result:
\bb
\lim_{|\lambda_k| \rightarrow \infty} \Big\|  \Big[ |\lambda_k| \alpha( -|\lambda_k| ) - 1 \Big] \ket{f^{n+1}} \Big\|   = 0
\label{eq: alphafinal}
\ee

\textbf{  $\delta$ term}
Let us repeat the steps in the light-front model,
\begin{eqnarray}
\nonumber
\Big\| (~ |\lambda_k| \Phi^{-1}(-|\lambda_k|) -1 ~ ) \ket{f^n} \Big\| &=& \Big\| ~  \Big[ |\lambda_k| (H_0 + |\lambda_k| + \mu_p)^{-1} (1 + K - U)^{-1}  ~ -1 \Big] \ket{f^n} \Big\|  \\
& \ &
\label{eq: gama}
\end{eqnarray}
Remember that in the limit $|\lambda_k| \rightarrow \infty$, $(1-(U-K))^{-1} $ can be expanded as Neumann series:
\bb
(1-(U-K))^{-1} +1 -1 = 1 + \sum\limits _{l=1} ^\infty (U-K)^l
\ee
and (\ref{eq: gama}) reduces to:
\begin{eqnarray}
\Big\| (~ |\lambda_k| \Phi^{-1}(-|\lambda_k|) -1 ~ ) \ket{f^n} \Big\| &=& \Big\| ~  \Big[ |\lambda_k| (H_0 + |\lambda_k| + \mu_p)^{-1} -1 \Big] \ket{f^n} 
+\Big[ |\lambda_k| (H_0 + |\lambda_k| + \mu_p)^{-1} \sum\limits _{l=1} ^\infty (U-K)^l \Big] \ket{f^n} \Big\|\nonumber
\end{eqnarray}
Using the triangle inequality:
\begin{eqnarray}
\Big\| (~ |\lambda_k| \Phi^{-1}(-|\lambda_k|) -1 ~ ) \ket{f^n} \Big\| &\leq & \Big\| ~  \Big[ |\lambda_k| (H_0 + |\lambda_k| + \mu_p)^{-1} -1 \Big] \ket{f^n} \Big\| \\
\nonumber
& \ & ~~~~~~~~~~~~+ \underbrace{\Big\| |\lambda_k| (H_0 + |\lambda_k| + \mu_p)^{-1} \Big\| } ~ \underbrace{ \Big\| \sum\limits _{l=1} ^\infty (U-K)^l \Big\| } \\
\nonumber 
& \ & ~~~~~~~~~~~~~~~~~~~~~~~~~~~ finite ~~~~~~~~~~~~~~~~~~~~ \rightarrow 0 ~~
\end{eqnarray}
where we take the limit $|\lambda_k| \rightarrow \infty$ in the second line and we deduce:
\bb
\lim_{|\lambda_k| \rightarrow \infty} \Big\| ~ |\lambda_k| \delta(-|\lambda_k|) -1 ~ \Big\| = \lim_{|\lambda_k| \rightarrow \infty} \Big\| ~ \Big[ |\lambda_k| ( H_0 + |\lambda_k| + \mu_p )^{-1} +1 \Big] \ket{f^n} ~  \Big\|
\label{eq:frcon}
\ee
Note that the resulting equation is (\ref{eq: decaycond}) in Theorem-1, $R(E)$ being the free resolvent. Since $\frac{1}{H_0 - E}$  is $indeed$ a $resolvent$, it must satisfy (\ref{eq: decaycond})  and we arrive at:
\bb
\lim_{|\lambda_k| \rightarrow \infty} \Big\| ~ \Big[ |\lambda_k| ( H_0 + |\lambda_k| + \mu_p )^{-1} -1 \Big] \ket{f^n} ~  \Big\| = 0
\label{eq: deltafinal}
\ee
Having shown that each term on the right hand side of the equations (\ref{eq: line1}) goes to zero as $|\lambda_k| \rightarrow \infty$, we  conclude that:
\bb
\lim_{|\lambda_k| \rightarrow \infty} \Big\| \Big[ |\lambda_k| R(-|\lambda_k|) -1 \Big] \begin{pmatrix}
\ket{f^{n+1}} \\
\ket{f^n}
\end{pmatrix}
\Big\| = 0
\ee
Therefore, we showed that $R(E) = \frac{1}{H-E}$  indeed defines a resolvent.

\subsubsection{ Common Domain of the Family $\Phi(E)$ on the Manifold}

We start by organizing the Principal Operator $\Phi(E)$ in the following way:
\begin{eqnarray}
\Phi(E) &=& \Big[ 1 + \sum_\sigma \frac{\lambda^2}{2\omega_\sigma} \frac{|f_\sigma|^2}{(H_0 - E + \omega_\sigma)} \frac{1}{(\omega_\sigma - \mu_p)}
\label{eq: firight} \\
\nonumber
& - & \sum_{\omega,\tau} \lambda^2 f_\sigma \frac{a_\sigma^\dagger}{\sqrt{2\omega_\sigma}} \frac{1}{H_0 - E + \omega_\sigma + \omega_\tau} \frac{1}{H_0 - E + \omega_\tau + \mu_p} \frac{a_\tau}{\sqrt{2\omega_\tau}} f_\tau \Big]  (H_0 - E + \mu_p)
\end{eqnarray}

Recall that we are working on a sector of the full Fock space, $ \mathcal{H} = \mathcal{F}^{(n+1)} \otimes \chi _\downarrow \oplus \mathcal{F}^{(n)} \otimes \chi_\uparrow $, which is a Hilbert space. Call the domain of $H_0$ as $D(H_0)$, which is dense in $\mathcal{F}^{(n)}$ for any $n$. Moreover, $H_0$ is closed on this domain being a self-adjoint operator.
Renaming the terms in \ref{eq: firight}, we rewrite $\Phi(E)$ as:
\bb
\Phi(E) = [ 1 + \mathcal{K} (E) + \mathcal{U} (E) ] ( H_0 - E + \mu_p )
\ee

To fix $D(H_0)$ to be the common domain of $\Phi(E)$, we want to show that $ \mathcal{K} (E) $ and $ \mathcal{U} (E) $ are bounded, $E$ being complex.  Since $\mathcal{K}(E)$ and $H_0$ commute, the new splitting of $\Phi(E)$ does not effect the bound we found for $K(E)$ previously in 1.3.2, which is:
\begin{eqnarray}
||\mathcal{K}(E)|| &\leq & \mathcal{C} ~\frac{\pi}{\sqrt{m}} \frac{1}{ \sqrt{(n+1)m - \operatorname{Re}(E)}} +  \frac{\mathcal{C}}{V(\mathcal{M})} \Big[ \frac{1}{2 (nm + \mu_p - \operatorname{Re}(E)) (m-\mu_p)^2} \Big]
\label{eq: Kbound2}
\end{eqnarray}
For $\mathcal{U}(E)$, we had previously worked with $E$ chosen on the real axis and the result must be generalized to the complex case. We start  by collecting the terms using Feynman parametrization:
\begin{eqnarray}
\nonumber
\mathcal{U}(E) &=& \sum_{\sigma,\tau} \lambda^2 f_\sigma \frac{a_\sigma^\dagger}{\sqrt{2\omega_\sigma}} \int_0 ^1 \frac{du}{[H_0 - E + (1-u) \mu_p  + u \omega_\sigma + \omega_\tau]^2 } \frac{a_\tau}{\sqrt{2\omega_\tau}} f_\tau \\
\nonumber
&=& \sum_{\sigma,\tau} \lambda^2 f_\sigma \frac{a_\sigma^\dagger}{\sqrt{2\omega_\sigma}} \int_0 ^1 du \int_0 ^\infty sds e^{-s(H_0 - E) -s \mu_p (1-u)}  e^{-su \omega_\sigma} e^{-s \omega_\tau} \frac{a_\tau}{\sqrt{2\omega_\tau}} f_\tau
\end{eqnarray}
Taking the norm:
\begin{eqnarray}
\nonumber
||\mathcal{U}(E)|| & \leq & \lambda^2 \int_0 ^\infty s ds \int _0 ^1 du  e^{-s\mu_p (1-u)} \Big| \Big| \sum_\sigma \frac{ a^\dagger _\sigma f_\sigma}{\sqrt{2\omega_\sigma}} e^ {-su\omega_\sigma} e^{sE}e^{-sH_0} \sum_\tau e^{-s\omega_\tau} \frac{a_\tau f_\tau}{\sqrt{2\omega_\tau}} \Big| \Big| \\
&\leq & \lambda^2 \int_0 ^\infty s ds \int _0 ^1 du  e^{-s\mu_p (1-u)} e^{s \operatorname{Re}(E)} \Big| \Big| \phi^{(-)} (f) e^{-sH_0} \phi^{(+)} (g) \Big| \Big|
\label{eq: Ucx}
\end{eqnarray}
where we define:
\begin{eqnarray}
\nonumber
\phi^{(-)} (f(s)) =  \sum_\sigma \frac{a^\dagger _\sigma f_\sigma(\bar{x})}{\sqrt{2\omega_\sigma}} e^ {-su\omega_\sigma}\quad  {\rm and} \quad   \phi^{(+)} (g(su)) =  \sum_\tau \frac{a _\tau f_\tau(\bar{x})}{\sqrt{2\omega_\tau}} e^ {-s\omega_\tau}
.\end{eqnarray}
Now we can estimate $e^{-sH_0}$ as $e^{-s(n-1)m}$,  we end  up with the following inequality:
\begin{eqnarray}
||\mathcal{U}(E)|| & \leq & \lambda^2 \int_0 ^\infty s ds \int _0 ^1 du  e^{s\mu_p u} e^{-s[(n-1)m+\mu_p-\operatorname{Re}(E)]} || \phi^{(-)}(g) \phi^{(+)}(f) ||
\label{eq: ue}
\end{eqnarray}
We need  to show that the integral is finite since the generalized heat kernels appearing in $\phi^{(-)}$ and $\phi^{(+)}$ are singular around $0^+$. We recall the estimate $ || \phi^{(-)} (f) \phi^{(+)}(g) ||\leq n||f||||g|| $ as discussed in the previous part. 
As a result we have,
\bb
||f||^2 = \sum_\sigma \frac{|f_\sigma (\bar{x})|^2}{2\omega_\sigma}e^{-2s \omega_\sigma} ~~ , ~~ ||g||^2 = \sum_\sigma \frac{|f_\sigma (\bar{x})|^2}{2\omega_\sigma}e^{-2us \omega_\sigma} 
\ee
We first estimate $||f||$ and $||g||$ by employing an  integral identity:
\begin{equation}
    \frac{e^{-2s\omega_\sigma}}{\omega_\sigma}={1\over \sqrt{\pi}} \int_0^\infty e^{-t(m^2+\sigma^2)-4s^2/t}{dt\over t^{1/2}}
.\end{equation},
then we have
\begin{eqnarray}
||f||^2 &=& C_1   \int \limits_0 ^\infty {dt\over t^{1/2}}  e^{-4s^2/t-m^2t}  \underbrace{\sum_\sigma e^{ -\sigma^2 t} ~ |f_\sigma (\bar{x})|^2} \\
\nonumber
&\ & ~~~~~~~~~~~~~~~~~~~~~~~~~~~~ K_{t} (\bar{x}, \bar{x}) \leq \frac{A}{t} + \frac{1}{V(\mathcal{M})}
\label{eq: f2}
\end{eqnarray}
We work with the most singular part:
\begin{eqnarray}
\nonumber
||f||_{sing}^2 &\leq & C_2\int \limits_0 ^\infty t^{-3/2} e^{-m^2t- 4s^2 / t} dt \leq  \frac{C}{s} e^{-2ms}
\end{eqnarray}
and similarly,
\bb
||g||_{sing} \leq \frac{C}{us} e^{-2mus}
\ee
We need  to substitute these estimates into (\ref{eq: ue}) to show that $||\mathcal{U}||$ is bounded:
\begin{eqnarray}
\nonumber
||\mathcal{U}(E)||_{sing} &\leq & C \lambda^2 n \int \limits_0 ^\infty s~ds \int \limits_0 ^1 du e^{s\mu_p u} e^{-s[(n-1)m+\mu_p-\operatorname{Re}(E)]} \frac{e^{-ms}}{\sqrt{s}} \frac{e^{-mus}}{\sqrt{su}} \\
\nonumber
&\leq & C \lambda^2 n \int \limits_0 ^\infty ds e^{-ms} \int \limits_0 ^1 du \frac{e^{-s(m-\mu_p)u}}{\sqrt{u}} e^{-s(nm+ \mu_p-\operatorname{Re}(E))} \\
\nonumber
&\leq & C \lambda^2 n \int \limits_0 ^\infty e^{-s[nm + \mu_p - \operatorname{Re}(E)]} ds \\
&\leq & C \lambda^2 n \frac{1}{nm + \mu_p - \operatorname{Re}(E)}
\end{eqnarray}
where we have used:
\bb
\int \limits_0 ^1 du \frac{e^{-s(m-\mu_p)u}}{\sqrt{u}} ~~\leq ~~ \int \limits_0 ^1 \frac{du}{\sqrt{u}} ~ = ~ 2
\ee
Since the most singular part is finite, the rest certainly is. This concludes that $\mathcal{K}(E)$ and $\mathcal{U}(E)$ are bounded, thus we can choose the domain of $H_0$ as the common domain of the family $\Phi (E)$ on the open domain $\Omega=\{ E \in \mathbf{C}| \operatorname{Re}(E)<nm+\mu_p\}$. 

As an essential ingredient of our arguments,  we need to establish  that $\Phi(E)$ is closed on its domain $D(\Phi(E)) = D = D(H_0)$, 
The argument is essentially algebraic, therefore the first part  is the same as in Section {\bf I.5.2}, so we do not repeat it,  in the same way,  for  $\operatorname{Re}(E) \leq \operatorname{Re}(E_*)$, we can show  $\Phi(E)$ is closed. Only the part where we show that the difference is a bounded operator needs to be discussed. For $\operatorname{Re}(E) > \operatorname{Re}(E_*)$, we rearrange according to (\ref{eq: phidif}) :
\begin{eqnarray}
 \nonumber
 \Phi(E) - \Phi(E_*) &=& T(E,E_*) (E_* - E) \\
 & =&  (E_* - E) \Big[ 1
  + \lambda^2 \sum_\sigma \frac{|f_\sigma | ^2}{2 \omega_\sigma} \frac{1}{(H_0-E+\omega_\sigma)(H_0-E_*+\omega_\sigma)} 
 \label{eq: e*}
\\
 \nonumber
 &+& \lambda^2  \sum_{\sigma,\tau} f_\sigma (\bar{x}) \frac{a_\sigma ^\dagger}{\sqrt{2\omega_\sigma}} \frac{1}{(H_0-E+\omega_\sigma+\omega_\tau)}
 \frac{1}{(H_0-E_*+\omega_\sigma+\omega_\tau)} \frac{a_\tau}{\sqrt{2\omega_\tau}} f_\tau (\bar{x}) \Big]
 \end{eqnarray}
 We want to show that $T(E,E_*)$ is bounded. Calling the second term in square brackets $A$, we proceed similar to previous calculations and  show that it is bounded. Again we apply Feynman parametrization followed by a subordination and take the norm. Estimating $H_0 > nm$ as well as recognizing the heat kernel as before and substituting the bound given in (\ref{eq: hkestimate}), we find:
 \begin{eqnarray}
\nonumber
 ||A|| \leq  \int_0 ^1 du_1du_2 du_3 \int_0 ^\infty s^3 ds e^{-s(nm-\operatorname{Re}(E)) u_2 } e^{-s(nm-\operatorname{Re}(E_*))u_3} 
  \int d \xi \frac{m e^{-s^2m^2/4\xi} e^{-\xi}} {\xi^{3/2}} \Big( \frac{C}{\xi/m^2} + \frac{1}{V(\mathcal{M})} \Big) 
 \end{eqnarray}
 We compute the most singular term (the first part) of the integral:
 \begin{eqnarray}
 \nonumber
 ||A||_{sing} &\leq & C \int ds du_1 du_2 du_3 ~ \delta(1- \sum_i u_i) ~ e^{-s(nm-\operatorname{Re}(E))u_2}   e^{-s(nm-\operatorname{Re}(E_*))u_3} ~ e^{-sm(u_1 + u_2 + u_3)} \\
 &\leq & C \int du_1 du_2 du_2 ~ \delta(1- \sum_i u_i)  \frac{1}{mu_1 + ((n+1)m - \operatorname{Re}(E))u_2 + ((n+1)m - \operatorname{Re}(E_*))u_3} \nonumber\\
 & \leq & C \int \limits_{0 \leq u_2 + u_3 \leq 1} \frac{1}{[(n+1)m-\operatorname{Re}(E)]u_2 + [(n+1)m-\operatorname{Re}(E_*)] u_3}
 \end{eqnarray}
 where in the last line we have ignored a positive term $m(1-u_2-u_3)$ in the denominator. Passing to polar coordinates:
 \begin{eqnarray}
 \nonumber
||A||_{sing} & \leq & C \int \limits_0 ^1 \int \limits _0 ^{\pi /2} \frac{\rho d\rho d\theta}{[(n+1)m-\operatorname{Re}(E)]\rho \cos\theta + [(n+1)m-Re(E_*)]\rho \sin\theta} \\
 & \leq & C \frac{1}{\sqrt{(n+1)-\operatorname{Re}(E)} \sqrt{(n+1)-\operatorname{Re}(E_*)}}
 \end{eqnarray}
 Note that we absorb every constant we encounter into $C$. If the most singular part is bounded, the other part certainly is. Therefore, we have shown that $A$ is bounded.\\
 We now show the boundedness of the third term in square brackets in (\ref{eq: e*}), call it $B$ for simplicity.

 \begin{eqnarray}
 \nonumber
 B & = & \lambda^2  \sum_{\sigma,\tau} f_\sigma (\bar{x}) \frac{a_\sigma ^\dagger}{\sqrt{2\omega_\sigma}} \frac{1}{(H_0-E+\omega_\sigma+\omega_\tau)(H_0-E_*+\omega_\sigma+\omega_\tau)} \frac{a_\tau}{\sqrt{2\omega_\tau}} f_\tau (\bar{x}) \\
\nonumber
 &=& \lambda^2 \sum_{\sigma,\tau} \int \limits_0 ^\infty sds \int \limits_0 ^1 du f_\sigma (\bar{x}) \frac{a_\sigma ^\dagger}{\sqrt{2\omega_\sigma}} e^{-s\omega_\sigma} e^{-sH_0} e^{s(Eu+E_*(1-u))} e^{-s\omega_\tau}  \frac{a_\tau}{\sqrt{2\omega_\tau}} f_\tau (\bar{x}) 
\end{eqnarray}
 Taking the norm and replacing $\operatorname{Re}(E_*)$ by $\operatorname{Re}(E)$ since $\operatorname{Re}(E_*) < \operatorname{Re}(E)$:
\begin{eqnarray}
 ||B|| &\leq& \lambda^2 \int \limits_0^ \infty s ds \int \limits_0 ^1 du  e^{s\operatorname{Re}(E)} || \phi^{(-)} (f) e^{-sH_0} \phi^{(+)} (f) ||
 \end{eqnarray}
we are faced with this estimate above, using the same notation $f$ as before and concentrating on the most singular part, we have an estimate:
 \begin{eqnarray}
 \nonumber
 ||B||_{(sing)} &\leq & n \lambda^2 \int \limits_0^ \infty sds \int \limits_0 ^1 du ~ e^{-s[(n-1)m-\operatorname{Re}(E)]} ||f||^2 \leq C ~ n  \int \limits_0^ \infty ds ~ e^{-s((n+1)m-\operatorname{Re}(E))}  \\
  & \leq & \frac{C ~ n}{(n+1)m-\operatorname{Re}(E)},  
\label{eq: Bbound}
\end{eqnarray}
which is finite. The other part is easier to estimate and it is finite. We repeat the argument given in Section {\bf I.5.2}, consequently, we conclude that $\Phi(E)$ is closed on its domain $D(\Phi(E)) =D(H_0)$.

\end{appendices}

\bibliography{ref.bib}{}

\begin{thebibliography}{10}

\bibitem{kaynak2009relativistic}
B.~T. Kaynak and O.~T. Turgut.
\newblock {The relativistic Lee model on Riemannian manifolds}.
\newblock {\em Journal of Physics A: Mathematical and Theoretical},
  42(22):225402, 2009.

\bibitem{lee1954some}
T.~D. Lee.
\newblock Some special examples in renormalizable field theory.
\newblock {\em Physical Review}, 95(5):1329, 1954.

\bibitem{pauli-kallen}
W.~Pauli and A.~O.~G. Kallen.
\newblock {On the Mathematical structure of T.D. Lee's model of a
  renormalizable field theory}.
\newblock {\em Kgl. Danske Videnskab. Selskab, Mat. Fys. (DOI:
  10.1007/978-3-319-00627-7-94, CERN-55-29)}, 30(7):1--23, 1955.

\bibitem{schweber2011introduction}
Silvan~S Schweber.
\newblock {\em An introduction to relativistic quantum field theory}.
\newblock Courier Corporation, 2011.

\bibitem{decelles1959}
P.~DeCelles and G.~Feldman.
\newblock {Dispersion Relations in the Lee model}.
\newblock {\em Nuclear Physics}, 14:517, 1959.

\bibitem{amado}
R.~D. Amado.
\newblock {$V-\theta$ Collisions in the Lee model}.
\newblock {\em Physical Review}, 122(5):696, 1961.

\bibitem{sommerfield1965}
C.~M. Sommerfield.
\newblock {Solution of the integral equation for V-$\theta$ scattering in the
  Lee model}.
\newblock {\em Journal of Math Phys.}, 6(7):1170, 1965.

\bibitem{bolsterli1968}
M.~Bolsterli.
\newblock {Algebraic Solution in the $V-\theta$ sector of the Lee model}.
\newblock {\em Physical Review}, 166(5):1760, 1968.

\bibitem{pagnamenta1965}
A.~Pagnamenta.
\newblock {Solution of the Kallen-Pauli Equation}.
\newblock {\em Journal of Mathematical Physics}, 6(6):955, 1955.

\bibitem{maxon1965}
S.~M. Maxon and R.~B. Curtis.
\newblock {LSZ formalism in the Lee model}.
\newblock {\em Physical Review}, 137(4B):996, 1965.

\bibitem{bolsterli1983}
M.~Bolsterli.
\newblock {Field Operator Decomposition in the Lee model}.
\newblock {\em Physical Review D}, 27(12):2940, 1983.

\bibitem{Fuda1982}
M.~G. Fuda.
\newblock Lee model and three-particle equations.
\newblock {\em Physical Review D}, 25(4):1972, 1982.

\bibitem{wilson1970model}
Kenneth~G Wilson.
\newblock Model of coupling-constant renormalization.
\newblock {\em Physical Review D}, 2(8):1438, 1970.

\bibitem{Fuda1990}
M.~G. Fuda.
\newblock {Poincare-invariant Lee model}.
\newblock {\em Physical Review D}, 41(2):534, 1990.

\bibitem{thirring-book}
W.~Thirring and E.~M. Henley.
\newblock {\em Elementary quantum field theory}.
\newblock McGraw-Hill Book Company, 1962.

\bibitem{rajeev1999bound}
Sarada~G Rajeev.
\newblock {Bound States in Models of Asymtotic Freedom}.
\newblock {\em arXiv preprint hep-th/9902025}, 1999.

\bibitem{teo-erman-dogan}
C.~Dogan, F.~Erman, and O.~Teoman Turgut.
\newblock {Existence of Hamiltonians for some singular interactions on
  manifolds}.
\newblock {\em Journal of Mathematical Physics}, 53(4):043511, 2012.

\bibitem{hirokawa1}
Asao Arai and Masao Hirokawa.
\newblock Stability of ground states in sectors and its application to the
  wigner--weisskopf model.
\newblock {\em Reviews in Mathematical Physics}, 13(04):513--528, 2001.

\bibitem{hirokawa2}
Masao Hirokawa.
\newblock Remarks on the ground state energy of the spin-boson model: An
  application of the wigner--weisskopf model.
\newblock {\em Reviews in Mathematical Physics}, 13(02):221--251, 2001.

\bibitem{facchi}
Paolo {Facchi}, Marilena {Ligab{\`o}}, and Davide {Lonigro}.
\newblock {Spectral properties of the singular Friedrichs-Lee Hamiltonian}.
\newblock {\em Journal of Mathematical Physics}, 62(3):032102, March 2021.

\bibitem{glimm-jaffe}
J~Glimm and A.~Jaffe.
\newblock {\em {Quantum Physics: A Functional Integral Point of View}}.
\newblock Springer-Verlag, NY, 2nd Edition, 1987.

\bibitem{turgut2018attractive}
O~Teoman Turgut and G{\"o}khan Yaln{\i}z.
\newblock An attractive $\phi^4$ theory in light-front coordinates.
\newblock {\em arXiv preprint arXiv:1812.09555}, 2018.

\bibitem{wilson}
Perry, Harindranath, and Wilson.
\newblock Light-front tamm-dancoff field theory.
\newblock {\em Physical review letters}, 65 24:2959--2962, 1990.

\bibitem{kato2013perturbation}
T.~Kato.
\newblock {\em Perturbation theory for linear operators}, volume 132.
\newblock Springer Science \& Business Media, 2013.

\bibitem{erman2014nondegeneracy}
Malkoc~B. Erman~F. and O.~T. Turgut.
\newblock {Nondegeneracy of the ground state for nonrelativistic Lee model}.
\newblock {\em Journal of Mathematical Physics}, 55(8):083522, 2014.

\bibitem{pazy2012semigroups}
Amnon Pazy.
\newblock {\em {Semigroups of linear operators and applications to partial
  differential equations}}, volume~44.
\newblock Springer Science \& Business Media, 2012.

\bibitem{everitt1997representation}
WN~Everitt, WK~Hayman, and G~Nasri~Roudsari.
\newblock On the representation of holomorphic functions by integrals.
\newblock {\em Applicable Analysis}, 65(1-2):95--102, 1997.

\bibitem{reed1975ii}
Michael Reed and Barry Simon.
\newblock {\em II: Fourier Analysis, Self-Adjointness}, volume~2.
\newblock Elsevier, 1975.

\bibitem{wust1972holomorphic}
Rainer W{\"u}st.
\newblock Holomorphic operator families and stability of self-adjointness.
\newblock {\em Mathematische Zeitschrift}, 125(4):349--358, 1972.

\bibitem{chavel1984eigenvalues}
Isaac Chavel.
\newblock {\em {Eigenvalues in Riemannian geometry}}, volume 115.
\newblock Academic press, 1984.

\bibitem{berline2003heat}
Nicole Berline, Ezra Getzler, and Michele Vergne.
\newblock {\em {Heat kernels and Dirac operators}}.
\newblock Springer Science \& Business Media, 2003.

\bibitem{wang1997global}
Jiaping Wang.
\newblock Global heat kernel estimates.
\newblock {\em pacific journal of mathematics}, 178(2):377--398, 1997.

\bibitem{rosenberg1997laplacian}
Steven Rosenberg.
\newblock {\em {The Laplacian on a Riemannian manifold: an introduction to
  analysis on manifolds}}, volume~31.
\newblock Cambridge University Press, 1997.

\bibitem{reed1978iv}
Michael Reed and Barry Simon.
\newblock {\em Methods of Modern Mathematical Physics: Analysis of Operators},
  volume~4.
\newblock Elsevier, 1978.

\end{thebibliography}
\bibliographystyle{unsrt}

\end{document}